\documentclass[twocolumn,footinbib,floatfix,showpacs,preprintnumbers,amsmath,amssymb,pra,superscriptaddress,longbibliography,notitlepage]{revtex4-1}

\usepackage{epsfig,amscd,wasysym}
\usepackage{times}
\usepackage{amsmath}
\usepackage{amsfonts}
\usepackage{amssymb}
\usepackage{graphicx}
\usepackage{color}
\usepackage{accents}
\usepackage[margin=20mm]{geometry}
\usepackage{array,mathtools,xcolor,physics,xr-hyper,hyperref,natbib}
\usepackage{subfigure}
\usepackage{dsfont}

\hypersetup{citecolor=violet}
\hypersetup{linkcolor=red}
\hypersetup{urlcolor=blue}
\hypersetup{colorlinks=true}

\newcommand{\be}{\begin{equation}}
\newcommand{\ee}{\end{equation}}
\newcommand{\bea}{\begin{eqnarray}}
\newcommand{\eea}{\end{eqnarray}}

\newcommand{\Ket}[1]{     |      #1  \,  \rrangle}
\newcommand{\BraKet}[2]{\llangle #1 | #2\rrangle}
\newcommand{\Bra}[1]{  \llangle #1    |} 
\newcommand{\phd}{\phantom\dagger}

\makeatletter 
\newsavebox{\@brx}
\newcommand{\llangle}[1][]{\savebox{\@brx}{\(\m@th{#1\langle}\)}%
  \mathopen{\copy\@brx\kern-0.5\wd\@brx\usebox{\@brx}}}
\newcommand{\rrangle}[1][]{\savebox{\@brx}{\(\m@th{#1\rangle}\)}%
  \mathclose{\copy\@brx\kern-0.5\wd\@brx\usebox{\@brx}}}
\makeatother

\newlength{\dhatheight} 

\newcommand{\qed}{\nobreak \ifvmode \relax \else
      \ifdim\lastskip<1.5em \hskip-\lastskip
      \hskip1.5em plus0em minus0.5em \fi \nobreak
      \vrule height0.75em width0.5em depth0.25em\fi}

\usepackage{enumitem}
\usepackage{lipsum}

\begin{document}

\title{Topological fingerprints in Liouvillian gaps}

	\author{K. Kavanagh\includegraphics[scale=0.066]{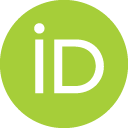}\href{https://orcid.org/0000-0002-6046-8495}}
	\email[Corresponding Author: ]{kvnkavanagh@gmail.com}
	\affiliation{Department of Physics, Faculty of Mathematics and Physics, University of Ljubljana, 1000 Ljubljana, Slovenia}
	\affiliation{School of Theoretical Physics, Dublin Institute for Advanced Studies, 10 Burlington Road, D04 C932, Ireland.}
	\affiliation{Department of Physics, Maynooth University, Maynooth, Co. Kildare, Ireland.}

	\author{J. K. Slingerland\includegraphics[scale=0.066]{./ORCID}\href{https://orcid.org/0000-0002-9112-269X}}
	\affiliation{Department of Physics, Maynooth University, Maynooth, Co. Kildare, Ireland.}
	\affiliation{School of Theoretical Physics, Dublin Institute for Advanced Studies, 10 Burlington Road, D04 C932, Ireland.}
	
	\author{S. Dooley\includegraphics[scale=0.066]{./ORCID}\href{https://orcid.org/0000-0002-2856-8840}}
	\affiliation{School of Theoretical Physics, Dublin Institute for Advanced Studies, 10 Burlington Road, D04 C932, Ireland.}
	\affiliation{Department of Physics, Trinity College Dublin, College Green, Dublin 2, Ireland}
	
	\author{G. Kells\includegraphics[scale=0.066]{./ORCID}\href{https://orcid.org/0000-0003-3008-8691}}
	\affiliation{Department of Physics, Maynooth University, Maynooth, Co. Kildare, Ireland.}
	\affiliation{School of Theoretical Physics, Dublin Institute for Advanced Studies, 10 Burlington Road, D04 C932, Ireland.}	

\preprint{DIAS-STP-24-01}
\date{\today}

\begin{abstract}
 Topology in many-body physics usually emerges as a feature of equilibrium quantum states. We show that topological fingerprints can also appear in the relaxation rates of open quantum systems. To demonstrate this we consider one of the simplest models that has two topologically distinct phases in its ground state: the Kitaev chain model for the $p$-wave superconductor. After introducing dissipation to this model we estimate the Liouvillian gap in both strong and weak dissipative limits. Our results show that a non-zero superconducting pairing opens a Liouvillian gap that remains open in the limit of infinite system size. At strong dissipation this gap is essentially unaffected by the topology of the underlying Hamiltonian ground state. In contrast, when dissipation is weak, the topological phase of the Hamiltonian ground state plays a crucial role in determining the character of the Liouvillian gap. We find, for example, that in the topological phase this gap is completely immune to changes in the chemical potential. On the other hand, in the non-topological phase the Liouvillian gap is suppressed by a large chemical potential. 
\end{abstract}

\pacs{74.78.Na, 74.20.Rp, 03.67.Lx, 73.63.Nm, 05.30.Ch}
\maketitle

\section{Introduction}
 
Topological condensed matter embraces the idea that band-structures can have non-trivial topologies and that these exotic forms can radically influence the behaviour of matter. This idea has been around for several decades and  has been enormously successful, with direct applications for metrology~\cite{Wen1990,Moore1991,DasSarma2005,Zhang2016}, spintronics~\cite{Qi2010TheQS}, and quantum information processing~\cite{Kitaev2003,Kitaev2006,NayakRev2008,Cheng2011}. The majority of works in this area focus on the equilibrium properties of matter, where topology most clearly arises in quantities calculated by integrating over momentum-space parametrizations of the single-particle excitation bands. Other indicators of topology (e.g. ground state degeneracies ~\cite{Hastings2005}, equivalences between local ground-state correlators~\cite{Levin2006, Coopmans2021}, bulk-boundary correspondences~\cite{Kitaev2012, Lee2016}, and tensor network classifications~\cite{Verstraete2017a, Verstraete2017b, Pollmann2021})  can be used beyond the implicitly non-interacting band theory of solids.

The fingerprints of topology are not however constrained to the equilibrated realm. Indeed, there has been much evidence of topology in recent years on numerous frontiers such as Floquet systems~\cite{Rieder2018, Fulga2019, Rudner2020, Simons2021}, non-Hermitian models~\cite{Kunst2018, Edvardsson2019, Bergholtz2021rev}, entanglement transitions in weakly measured models~\cite{Gebhart2020, Wang2021obs, Lavasani2021,Kells2022}, along with proposals to engineer topological steady states in open quantum systems~\cite{Bardyn2013, Budich2015, Sieberer2016, Iemini2016, Goldman2016, Barbarino2020, Tonielli2020, Wolff2020, Mathey2020, Altland2021, Huang2021}.

In this paper, we ask whether coherent topological quantum effects can affect the relaxation behaviour of a system. We do this by focusing on the example of the Kitaev chain~\cite{Kitaev2003} (i.e., the spin-1/2 transverse XY model) in the presence of local bulk dephasing. Our first key result is that $p$-wave pairing is directly responsible for a Liouvillian gap that scales quadratically with the pairing strength, $\Delta$, and inversely to the dissipative rate, $\epsilon$. Moreover, the gap remains open even as the system size $N\rightarrow\infty$. This result applies where the strength of dissipation, $\epsilon$, is far greater than any component of the Hamiltonian. We arrive at this result analytically via a perturbative mapping of the whole model to an XXZ chain. Crucially we see that, although the anisotropy massively boosts the relaxation rate, the details of underlying quantum order are largely inconsequential.    

This is in contrast to the weakly dissipative regime, where we show that this picture is effectively reversed. Here, although the gap still remains constant with respect to system size, it now effectively scales linearly with the product of  $\Delta \epsilon$. Crucially, in this regime, we also see that the underlying order of the Hamiltonian can have a striking effect on the system's relaxation rates. For example when the underlying Hamiltonian is in a topological phase, the Liouvillian gap evaluates to be independent of the chemical potential $\mu$. Outside of this region, the gap is parametrically reduced at a rate that tends to  $1/ \sqrt{\mu}$ for large $\mu$. 

We obtain these results in the framework of ``operator quantization''~\cite{Prosen2008, Prosen2008b}, making use in particular of the structure of the Liouvillian superoperator in the canonical matrix representation~\cite {Kells2015b, Kells2018, Kavanagh2022}. In the strongly dissipative regime this allows us to directly identify the relevant effective subspace in the kernel of the dissipator, and within that subspace construct an effective perturbation theory that reveals the mapping to the XXZ model. In the weakly dissipative limit, this same representation enables projections of the dissipative terms into the kernel of the Hamiltonian commutator and from this, to calculate an approximation of the Liouvillian gap. By extrapolating to the thermodynamic limit we can then show that the resulting evaluation for the gap is distinctly topological in character.  

The methodology we use can be applied for a number of dissipative processes but is particularly transparent in cases where the dissipative jump operators are Hermitian. We show how this works with the example of bulk dephasing, and we provide an additional example for the Hermitian formulation of the Symmetric Simple Exclusion Process (SSEP)~\cite{Eisler2011} in appendix~\ref{app:Oexamples}. We argue however that it can also give insight into other processes that do not have Hermitian jump operator formulations, showing in particular how it also explains the observed behaviour of the TXY-TASEP system~\cite{Kavanagh2022}. A similar methodology was used to study a dissipative quantum compass model Ref.~\cite{Shibata2019a} and the dissipative quantum Ising chain~\cite{Shibata2019b}. Here similar sharp transitions in the gap behaviour are also observed. Other atypical relaxation behaviours have been found in Sachdev-Ye-Kitaev (SYK) models subject to dissipation~\cite{Sa2022,Sa2023}. Therein, treating the system in the Keldysh formalism one can find a Liouvillian gap dependent only on the fixed ratio of number of Majorana and Lindblad jump operators, Hamiltonian coupling and dissipation strength. Moreover, a rich variety of anomalous relaxation rates can be found in open quanutm circuits~\cite{Yoshimura2023}. In this context, both exponential and non-exponential behaviours of the relaxation rate may be obtained.

\section{Model and symmetries}

\begin{figure}[t]
	\centering
	\includegraphics[width=1\columnwidth, page=6]{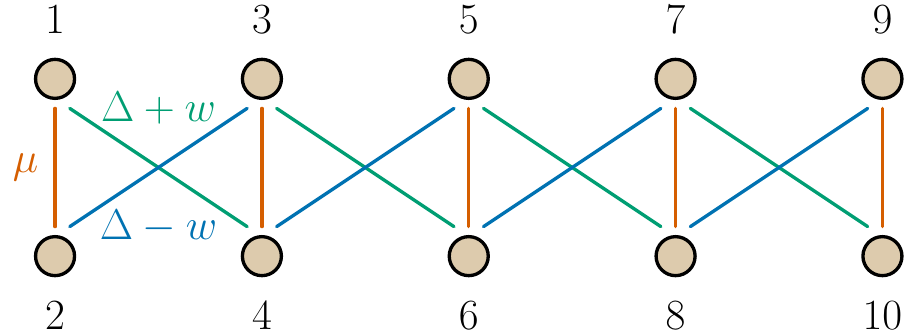}
	\caption{(Color Online) The Kitaev chain Hamiltonian Eq.~\ref{eq:H} can be visualised as a ladder with odd Majorana operators on one leg and even Majorana operators on the opposite leg. The representation of the Liouvillian in terms of the creation and annihilation superoperators in Eq.~\ref{eq:mathcal_H} can similarly be visualised as a ladder with Majorana operators that can hop across the bonds.}
\label{fig:ladder}
\end{figure}

\subsection{Hamiltonian and Lindblad master equation}

We consider a system of $2N$ Majorana fermions $\{ \hat{\gamma}_{j} \}_{j=1}^{2N}$ with the anti-commutation relations $\{ \hat{\gamma}_{j} , \hat{\gamma}_{j'} \} = 2\delta_{j,j'}\hat{\mathbb{I}}$ and the quadratic Hamiltonian (see Fig. \ref{fig:ladder})
 \begin{eqnarray} 
 \hat{H} = - \frac{i\mu}{2} \sum_{n=1}^N \hat{\gamma}_{2n-1}\hat{\gamma}_{2n} &+& \frac{i}{2} \sum_{n=1}^{N-1} [(\Delta+w)\hat{\gamma}_{2n-1}\hat{\gamma}_{2n+2} \nonumber \\ 
 &+& (\Delta-w)\hat{\gamma}_{2n}\hat{\gamma}_{2n+1} ] . \label{eq:H} 
 \end{eqnarray} 
The Hamiltonian may also be written in two equivalent forms that are perhaps more familiar to some readers. In terms of the Dirac fermion creation and annihilation operators $\hat{a}_n = (\hat{\gamma}_{2n-1} +i\hat{\gamma}_{2n})/2$ and $\hat{a}_n^\dagger = (\hat{\gamma}_{2n-1} - i\hat{\gamma}_{2n})/2$, the Hamiltonian $\hat{H}$ is the Kitaev model for a $p$-wave superconducting chain of $N$ sites~\cite{Kitaev2001}. In this case, $\mu$ is the chemical potential, $w$ is hopping strength, and $\Delta$ is the superconducting pairing strength. Alternatively, after a Jordan-Wigner transformation the Hamiltonian $\hat{H}$ is the quantum $XY$ model for a chain of $N$ spin-1/2 particles, with $\mu$ interpreted as magnetic field strength and $\Delta$ as the $XY$ anisotropy. 
  
  An important property of the Hamiltonian in Eq.~\ref{eq:H} is that its ground state has a symmetry-protected topological phase transition at $|\mu| = 2w$ (if $\Delta \neq 0$) \cite{Kitaev2001}. Another important property of Eq.~\ref{eq:H}, and of quadratic Hamiltonians in general, is that it can be easily reduced to its normal form $\hat{H} = \frac{1}{2}\sum_{n=1}^N \omega_n (\hat{\beta}^\dagger_n \hat{\beta}_n - \hat{\beta}_n \hat{\beta}_n^\dagger)$ via a Bogoliubov transformation,
\begin{equation} 
  \hat{\beta}_{n} = \sum_{j=1}^{2N} W_{nj}^{*} \hat{\gamma}_{j}.\label{eq:bogoliubov} 
\end{equation} 
Here, $\hat{\beta}_n^\dagger$ and $\hat{\beta}_n$ are Dirac fermion creation and annihilation operators obeying the anti-commutation relations $\{ \hat{\beta}_n^\dagger,  \hat{\beta}_{n'} \} = \delta_{n,n'}$, $\; n,n' \in \{ 1,\hdots,N \}$.

To model an interaction with an external environment, we suppose that the quantum state $\hat{\rho}$ evolves by the Lindblad master equation~\cite{Gorini1976, Lindblad1976}
\begin{equation} 
\frac{d\hat{\rho}}{dt} = - i\mathbb{H}(\hat{\rho}) + \mathbb{D}(\hat{\rho}) \equiv \mathbb{L}(\hat{\rho}) , \label{eq:GKSL} 
\end{equation} 
which is comprised of two parts, the Hamiltonian commutator $\mathbb{H}(\rho) = [\hat{H}, \rho]$, and the dissipator
\begin{equation} 
\mathbb{D}(\rho) = \sum_n ( \hat{\ell}_n \hat{\rho} \hat{\ell}_n^\dagger -\frac{1}{2} \hat{\ell}_n^\dagger\hat{\ell}_n \hat{\rho} - \frac{1}{2}\hat{\rho}\hat{\ell}_n^\dagger \hat{\ell}_n ) . \label{eq:dissipator}
\end{equation}
We choose Lindblad jump operators of the form 
\begin{equation} 
	\hat{\ell}_{n} = - i \sqrt{\epsilon} \hat{\gamma}_{2n-1}\hat{\gamma}_{2n} , \qquad n \in \{ 1,2,\hdots, N \}. \label{eq:jump_ops}
\end{equation} 
In terms of the Dirac fermions $\hat{a}_n$ these Lindblad operators are $\hat{\ell}_n = \frac{1}{4}\sqrt{\epsilon} (\hat{a}_n \hat{a}^{\dagger}_{n} - \hat{a}^\dagger_n \hat{a}_n)$. Alternatively, after a Jordan-Wigner transformation, the Lindblad operators in the spin-1/2 picture are $\hat{\ell}_n = \sqrt{\epsilon}\hat{\sigma}_{n}^{z}$, representing local qubit dephasing.

The combination of the commutator and the dissipator in Eq. \ref{eq:GKSL} is referred to as the Liouvillian $\mathbb{L}$. We note that bold symbols $\mathbb{H}$, $\mathbb{D}$, $\mathbb{L}$ represent superoperators, i.e., linear maps that take operators to operators.

Dephasing has wide application in non-stationary dynamics~\cite{Buca2019}, environment assisted procesess~\cite{Sarovar2008,Rebentrost2009,Sarkar2020} and discrete time crystals~\cite{Sarkar2022a,Sarkar2022b}. In the context of this paper, dephasing is acts an illustrative example of a class of dissipative processes for which our methodology can be applied.

\subsection{Matrix representation of the Liouvillian superoperator}

We introduce strings of Majorana operators defined as 
\begin{equation} 
\hat{\Gamma}^{\vec{\nu}} = \hat{\gamma}_{1}^{\nu_{1}} \hat{\gamma}_{2}^{\nu_{2}} \dots \hat{\gamma}_{2N}^{\nu_{2N}} , 
\end{equation} 
where the bitstring $\vec{\nu}$ indicates which $\hat{\gamma}_j$ are present ($\nu_j = 1$) or absent ($\nu_j = 0$) in a given operator string $\hat{\Gamma}^{\vec{\nu}}$. Using the Hilbert-Schmidt inner product $\text{Tr} (\hat{A}^\dagger \hat{B})$ between any two operators $\hat{A}$ and $\hat{B}$, it is straightforward to show that the operators $\hat{\Gamma}^{\vec{\nu}}/\sqrt{2^N}$ are orthonormal, i.e., $\text{Tr}[(\hat{\Gamma}^{\vec{\nu}})^\dagger \hat{\Gamma}^{\vec{\nu}'}]/2^N = \delta^{\vec{\nu}, \vec{\nu}'}$. The $4^N$ operators $\hat{\Gamma}^{\vec{\nu}}/\sqrt{2^N}$ therefore form an orthonormal basis for the $4^N$-dimensional space of superoperators.

The Majorana operator basis $\{ \hat{\Gamma}^{\vec{\nu}} \}$ allows us to represent superoperators in a convenient matrix form. For example, the Liouvillian superoperator $\mathbb{L}$ can be represented as a matrix $\mathcal{L}$ with the elements 
\begin{equation} 
\mathcal{L}^{\vec{\nu},\vec{\nu}'} = \text{Tr}[(\hat{\Gamma}^{\vec{\nu}})^\dagger\mathbb{L}(\hat{\Gamma}^{\vec{\nu}'})] . \label{eq:L_elements} 
\end{equation} 
Another way of thinking about this is to vectorize the quantum state $\hat{\rho} \to |\rho\rrangle$, e.g., by stacking the rows of the $2^N \times 2^N$ matrix $\hat{\rho}$ to make a single $4^N$-dimensional vector $|\rho\rrangle$. The action of a superoperator on $\hat{\rho}$ is then a matrix multiplication applied to the vector $|\rho\rrangle$. For example, the Liouvillian superoperator becomes 
\begin{equation} 
\mathbb{L}(\hat{\rho}) \to \mathcal{L}|\rho\rrangle = (\mathcal{H} + \mathcal{D})|\rho\rrangle , 
\end{equation} 
where 
\begin{equation}
	\begin{split}
		\mathcal{H} &= -i(\hat{H}\otimes\hat{\mathbb{I}}-\hat{\mathbb{I}}\otimes \hat{H}^{T}), \\ 
		\mathcal{D} &=  \sum_{n} ( \hat{\ell}_n^* \otimes \hat{\ell}_n - \frac{1}{2} \hat{\mathbb{I}}\otimes \hat{\ell}_{n}^\dagger \hat{\ell}_{n} - \frac{1}{2} \hat{\ell}_{n}^{T} \hat{\ell}_{n}^{*} \otimes \hat{\mathbb{I}} ), 
	\end{split}
\end{equation} 
are the matrix representations of the superoperators $\mathbb{H}$ and $\mathbb{D}$, respectively. In this framework, the matrix elements of $\mathcal{L}$ in the Majorana operator basis Eq.~\ref{eq:L_elements} are equivalently written as 
\begin{equation} 
\mathcal{L}^{\vec{\nu},\vec{\nu}'} = \llangle \Gamma^{\vec{\nu}} |  \mathcal{L} | \Gamma^{\vec{\nu}'} \rrangle . \label{eq:Lcanonical} 
\end{equation}

Moreover, following Prosen~\cite{Prosen2008}, we view the vectorized operator basis $|\Gamma^{\vec{\nu}}\rrangle$ as a set of Fock ``states'', with occupation of ``operator modes'' given by the indices $\vec{\nu}$. Analagous to the usual fermionic Fock states, we can then define creation and annihilation superoperators
 \begin{eqnarray}
 	\mathcal{G}_j^\dagger |\Gamma^{\vec{\nu}}\rrangle &=& \delta_{\nu_j, 0} | \gamma_j \Gamma^{\vec{\nu}} \rrangle = \left\{ \begin{array}{lr} | \gamma_j \Gamma^{\vec{\nu}} \rrangle, & \text{if } \nu_j = 0 \\
 	 0 , & \text{if } \nu_j = 1  \end{array} \right. , \label{eq:supercreation} \\ 
 	\mathcal{G}_j |\Gamma^{\vec{\nu}}\rrangle &=& \delta_{\nu_j, 1} | \gamma_j \Gamma^{\vec{\nu}} \rrangle = \left\{ \begin{array}{lr} 0 , & \text{if } \nu_j = 0 \\ 
 			| \gamma_j \Gamma^{\vec{\nu}} \rrangle , & \text{if } \nu_j = 1  \end{array} \right. , \label{eq:superdestroy} 
 \end{eqnarray}
 which obey the usual fermionic anti-commutation relations $\{ \mathcal{G}_j, \mathcal{G}_{j'}^\dagger \} = \delta_{j,j'}$ and $\{ \mathcal{G}_j, \mathcal{G}_{j'} \} = \{ \mathcal{G}_j^\dagger, \mathcal{G}_{j'}^\dagger \} = 0$.
 We also define the number superoperator
 \begin{equation} 
 	\mathcal{N} = \sum_{j=1}^{2N} \mathcal{G}_j^\dagger \mathcal{G}_j , \label{eq:num_spop}
 \end{equation} 
 which has the property that $\mathcal{N} | \Gamma^{\vec{\nu}}\rrangle = |\vec{\nu}|^2 | \Gamma^{\vec{\nu}} \rrangle$, i.e., the Majorana strings are eigenoperators of $\mathcal{N}$ and the eigenvalue is the number of Majorana operators appearing in $|\Gamma^{\vec{\nu}}\rrangle = | \hat{\gamma}_{1}^{\nu_{1}} \hat{\gamma}_{2}^{\nu_{2}} \dots \hat{\gamma}_{2N}^{\nu_{2N}} \rrangle$.

 We note that the creation and annihilation superoperators in Eqs.~\ref{eq:supercreation},~\ref{eq:superdestroy} are defined via multiplication by the created/annihilated operator $\gamma_j$ from the left, i.e., $| \gamma_j \Gamma^{\vec{\nu}} \rrangle$. However, we can also show, using the commutation relations $\{ \hat{\gamma}_j , \hat{\gamma}_{j'} \} = 2\delta_{j,j'}\hat{\mathbb{I}}$, that this is equivalent to 
 \begin{eqnarray}
 	\mathcal{G}_j^\dagger |\Gamma^{\vec{\nu}}\rrangle &=& - \delta_{\nu_j, 0} (-1)^{\mathcal{N}}| \Gamma^{\vec{\nu}}  \gamma_j \rrangle, \label{eq:supercreation_R} \\ 
 	\mathcal{G}_j |\Gamma^{\vec{\nu}}\rrangle &=& \delta_{\nu_j, 1} (-1)^\mathcal{N} | \Gamma^{\vec{\nu}} \gamma_j \rrangle, \label{eq:superdestroy_R}
 \end{eqnarray} 
 where the created/annihilated operator now multiplies from the right. It follows from Eqs.~\ref{eq:supercreation},~\ref{eq:superdestroy} that $| \gamma_j \Gamma^{\vec{\nu}} \rrangle = (\mathcal{G}_j + \mathcal{G}_j^\dagger) | \Gamma^{\vec{\nu}} \rrangle$ and it follows from Eqs.~\ref{eq:supercreation_R},~\ref{eq:superdestroy_R} that $| \Gamma^{\vec{\nu}} \gamma_j \rrangle = (-1)^\mathcal{N} (\mathcal{G}_j - \mathcal{G}_j^\dagger) | \Gamma^{\vec{\nu}} \rrangle$, so that:
 \begin{eqnarray} 
 	| \gamma_j \rho \rrangle &=& (\mathcal{G}_j + \mathcal{G}_j^\dagger) | \rho \rrangle, \label{eq:left_mult} \\ 
 	| \rho \gamma_j \rrangle &=& (-1)^\mathcal{N} (\mathcal{G}_j - \mathcal{G}_j^\dagger) | \rho \rrangle. \label{eq:right_mult}
 \end{eqnarray} 
These equations can be used to rewrite the Lindblad master equation in the superoperator matrix representation (in terms of the superoperator creation and annihilation operators). A short calculation shows that the Hamiltonian commutator $-i\mathbb{H}(\rho) = -i [\hat{H}, \rho] \to \mathcal{H} |\rho\rrangle$ is expressed as 
\begin{equation}
	\begin{split}
		\mathcal{H} &= -\mu \sum_{n=1}^N (\mathcal{G}_{2n-1}\mathcal{G}_{2n}^\dagger + \mathcal{G}_{2n-1}^\dagger \mathcal{G}_{2n}) \\ 
					&+ (\Delta + w)\sum_{n=1}^{N-1} (\mathcal{G}_{2n-1}\mathcal{G}_{2n+2}^\dagger + \mathcal{G}_{2n-1}^\dagger \mathcal{G}_{2n+2}) \\ 
					&+ (\Delta - w)\sum_{n=1}^{N-1} (\mathcal{G}_{2n}\mathcal{G}_{2n+1}^\dagger + \mathcal{G}_{2n}^\dagger \mathcal{G}_{2n+1}) . \label{eq:mathcal_H} 
	\end{split}
\end{equation}
Similar to the original Hamiltonian $H$ in Eq. \ref{eq:H}, this effective Hamiltonian in the superoperator picture is conveniently visualised as Majorana operators hopping on a two-leg ladder with sites of odd and even indices on opposite legs (see Fig.~\ref{fig:ladder}). The dissipator given in Eqs.~\ref{eq:dissipator},~\ref{eq:jump_ops}, transformed to the superoperator matrix representation, $\mathbb{D} \to \mathcal{D}$, is expressed as
\begin{equation}
  \mathcal{D} = \epsilon \sum_{n = 1}^{N} \left( [\mathcal{G}_{2n-1}^{\dagger}, \mathcal{G}_{2n-1}] [\mathcal{G}_{2n}^\dagger, \mathcal{G}_{2n}]  - \mathbb{I} \right) . \label{eq:mathcal_D}
\end{equation} The total Liouvillian superoperator is the sum $\mathcal{L} = \mathcal{H} + \mathcal{D}$.
 
\subsection{Symmetries of the Liouvillian superoperator}
\label{sec:block_structure}

The matrix representation of the superoperator $\mathcal{L} = \mathcal{H} + \mathcal{D}$ provides a convenient framework to identify its symmetries. For example, it commutes with the number superoperator $\mathcal{N}$ given in Eq. \ref{eq:num_spop}, i.e., $[\mathcal{L}, \mathcal{N}] = 0$. This means that the matrix $\mathcal{L}$ has a block structure in the Majorana operator basis, with blocks labelled by the eigenvalues $|\vec{\nu}|^2$ of $\mathcal{N}$. We denote the block corresponding to the $|\vec{\nu}|^2$ eigenspace as $\mathcal{L}_{|\vec{\nu}|^2}$. The superoperator matrices $\mathcal{H}$ and $\mathcal{D}$ also individually commute with $\mathcal{N}$ and have blocks denoted $\mathcal{H}_{|\vec{\nu}|^2}$ and $\mathcal{D}_{|\vec{\nu}|^2}$, respectively.

Another symmetry of our Liouvillian $\mathcal{L}$ is related to the superoperator defined as
\begin{equation} \mathcal{P} \equiv (\mathcal{G}_1 + \mathcal{G}_1^\dagger)(\mathcal{G}_2 + \mathcal{G}_2^\dagger) \hdots (\mathcal{G}_{2N} + \mathcal{G}_{2N}^\dagger) . 
\end{equation} 
This superoperator has the properties that $\mathcal{P} \mathcal{G}_j \mathcal{P}^{-1} = - \mathcal{G}_j^\dagger$ and $\mathcal{P} \mathcal{G}_j^\dagger \mathcal{P}^{-1} = - \mathcal{G}_j$ (i.e., it implements the transformation $\mathcal{G}_j \to -\mathcal{G}_j^\dagger$ and $\mathcal{G}_j^\dagger \to -\mathcal{G}_j$). It is not difficult to see that our Liouvillian $\mathcal{L}$ is invariant under this transformation, i.e., $\mathcal{P}\mathcal{L}\mathcal{P}^{-1} = \mathcal{L} \implies [\mathcal{L}, \mathcal{P}] = 0$. Despite first appearances, this symmetry has a fairly simple physical interpretation: since $\mathcal{P}^2 = (-1)^N$ its eigenvalues are $\pm i^N$. It turns out that the $+$ ($-$) eigenvalue is obtained when the superoperator $\mathcal{P}$ acts on density matrices $\Ket{\rho}$ with exclusively positive (negative) number parity. This is because $\mathcal{P}\Ket{\rho} = \Ket{\Gamma^{\vec{\nu} = \vec{1}}\rho}$ (from from Eq.~\ref{eq:left_mult}), where $\Gamma^{\vec{\nu} = \vec{1}} = (-i)^N\hat{\sigma}_z^{\otimes N}$ is proportional to the state number parity operator. The conservation of $\mathcal{P}$ by the Liouvillian $\mathcal{L}$ therefore corresponds to the conservation of the state number parity $\hat{\sigma}_z^{\otimes N}$ during the dynamics. 

Another consequence of the symmetry $\mathcal{P}\mathcal{L}\mathcal{P}^{-1} = \mathcal{L}$ is that any eigenoperator $\mathcal{L}\Ket{\alpha} = \lambda_\alpha \Ket{\alpha}$ of the Liouvillian also has the eigenoperator $\mathcal{P}\Ket{\alpha}$ with the same eigenvalue $\lambda_\alpha$. If $\mathcal{P}\Ket{\alpha} \neq \Ket{\alpha}$ then the corresponding eigenvalue is degenerate. Also, if the eigenoperator $\Ket{\alpha}$ is in the $\mathcal{N} = |\vec{\nu}|^2$ symmetry sector of the number superoperator then its partner eigenoperator $\mathcal{P}\Ket{\alpha}$ is in the $\mathcal{N} = 2N - |\vec{\nu}|^2$ symmetry sector. However, $\mathcal{N}$ and $\mathcal{P}$ do not commute, so $\mathcal{L}$ is not diagonalizable in both eigenbases simultaneously.

\section{Relaxation rates in the strong and weak dissipative limits}

It is straightforward to verify that for Hermitian jump operators the maximally mixed state $\hat{\rho}_\infty = \hat{\mathbb{I}}/2^{N}$ is a steady state of our Lindblad master equation, i.e., $\mathcal{L}\Ket{\mathbb{I}} = 0$. However, from the previous section, we see that when parity symmetry is present we can also expect that the state $\mathcal{P}\Ket{\Gamma^{\vec{0}}} = \Ket{\Gamma^{\vec{1}}}$ is also a steady state, $\mathcal{L} \Ket{\Gamma^{\vec{1}}} = 0$. Steady states with well-defined parity will therefore come in the form $\Ket{\hat{\rho}_\infty} = (\Ket{\hat{\Gamma}^{\vec{0}}} \pm \Ket{\hat{\Gamma}^{\vec{1}}})/2^N$.

Computing relaxation rates (inferred from the size of the Liouvillian gap) can be difficult for large systems. In the following section, we study this analytically by making approximations in two parameter regimes of interest: the strong and weak dissipative limits, i.e., the limits where the Hamiltonian component of the dynamics is weak or strong compared to the dissipative process.

The key feature we want to address is the role that Cooper pair creation and annihilation $\Delta$ plays in the steady-state relaxation rate.  In cases where the eventual steady-state is, up to parity considerations, a featureless infinite-temperature thermal state, there should be a significant effect that can be argued heuristically:  Cooper pair creation and annihilation will drive the system towards half-filling and the infinite temperature steady-state is naturally dominated by such half-filled states. One might reasonably expect then that Cooper pair creation and annihilation might help reduce the time it takes to reach this infinite temperature state. 

\subsection{Strong dissipative limit - Projection to the kernel of $\mathcal{D}$}

We begin by considering the limit where the Hamiltonian term $\mathcal{H}$ is a small perturbation to the dissipator term $\mathcal{D}$ (i.e., $\epsilon \gg w,|\Delta|,|\mu|$), the strong dissipative limit. In this limit, it is possible to construct an expansion of the Liouvillian  $\mathcal{L}_{\text{eff}} = \mathcal{L}^{(0)} + \mathcal{L}^{(1)} + \mathcal{L}^{(2)} + \dots$ where $\mathcal{L}^{(n)}$ are the $n^{\text{th}}$ order perturbations of the degenerate kernel  of $\mathcal{D}$. 

The kernel of the dissipator  (i.e. the set of states $\Ket{\hat{Z}_i}$ that are annihilated by $ \mathcal{D}$) are those states in which all Majorana basis operators $\Ket{\hat{\Gamma}^{\vec{\nu}}}$ have the property that $\nu_{2n-1} = \nu_{2n},\forall\, n$ (which can be verified by applying Eq. \ref{eq:mathcal_D} to such states and using properties Eq. \ref{eq:supercreation},~\ref{eq:superdestroy}). We note that the operators $\Ket{\hat{\Gamma}^{\vec{\nu}}}$ with $\nu_{2n-1} = \nu_{2n}$ can be visualized using Fig.~\ref{fig:ladder}: they are Majorana strings for which all rungs of the ladder are either occupied by Majoranas ($\nu_{2n-1} = \nu_{2n} = 1$) or unoccupied ($\nu_{2n-1} = \nu_{2n} = 0$).  States $\Ket{\hat{\Gamma}^{\vec{\nu}}}$ that do not have matching occupation numbers on either side of the ladder rung, come with an energy penalty of $2\epsilon $ for each unmatched rung~\footnote{Alternatively, switching to the spin picture through a Jordan-Wigner transformation, the operator $\hat{\Gamma}^{\vec{\nu}}$ with $\nu_{2n-1} = \nu_{2n}$ is simply a tensor product of single-qubit identity operators $\hat{\mathbb{I}}_n$ (if $\nu_{2n-1} = \nu_{2n} = 0$) or Pauli operators $\hat{\sigma}_n^z$ (if $\nu_{2n-1} = \nu_{2n} = 1$) at each spin site, which is clearly annihilated by the pure dephasing dissipator $\mathbb{D}(\hat{\Gamma}^{\vec{\nu}}) = \epsilon \sum_n (\hat{\sigma}_n^z \hat{\Gamma}^{\vec{\nu}} \hat{\sigma}_n^z - \hat{\Gamma}^{\vec{\nu}}) = 0$.}.

Defining the kernel projector as $ P = \sum_{i} \Ket{\hat{Z}_{i}} \Bra{\hat{Z}_{i}}$ and $Q = \mathbb{I}-P$,  we formally write out the effective Liouvillian terms to third order as~\cite{kato1949, bloch1958a, bloch1958b, messiah1962, lowdin1962, soliverez1969, rieder2012, moran2017}
\begin{equation}
	\begin{split}
		\mathcal{L}^{(0)} &= P \mathcal{D} P, \\
		\mathcal{L}^{(1)} &= P \mathcal{H} P, \\
		\mathcal{L}^{(2)} &= P \mathcal{H} \frac{Q}{ \mathcal{D}} \mathcal{H} P, \\
		\mathcal{L}^{(3)} &= P \mathcal{H} \frac{Q}{ \mathcal{D}} \mathcal{H}  \frac{Q}{ \mathcal{D}} \mathcal{H} P \\
  &\quad- \frac{1}{2} (P \mathcal{H} \frac{Q}{  \mathcal{D}^2} \mathcal{H}  P \mathcal{H} P +P \mathcal{H} P \mathcal{H} \frac{Q}{  \mathcal{D}^2} \mathcal{H}  P ).
	\end{split}
\end{equation}
The expression $\mathcal{L}^{(0)}$ is easily seen to vanish because $P$ projects to the kernel of $\mathcal{D}$. The situation is similar for $\mathcal{L}^{(1)}$ because $\mathcal{H}$ takes us completely out of the kernel  so that $\Bra{\hat{Z}_i} \mathcal{H} \Ket{\hat{Z}_j} =0 $  (see also Appendix \ref{app:eff_XXZ_deriv}).  Similar considerations also apply to $\mathcal{L}^{(3)}$ although to see why the first term vanishes requires a bit more thought.  

At second order within the degenerate subspace, the matrix $\mathcal{L}^{(2)}$ does not vanish and can be represented by (see App.~\ref{app:eff_XXZ_deriv})
\begin{equation}
		\mathcal{L}^{(2)} = J \sum_{n}\left[\left( \hat{\tau}_{n}^{x} \hat{\tau}_{n+1}^{x} + \hat{\tau}_{n}^{y} \hat{\tau}_{n+1}^{y}\right) +\Delta_{a} \left(\hat{\tau}_{n}^{z}\hat{\tau}_{n+1}^{z} - \hat{\mathbb{I}}\right)\right],\label{eq:effective_XXZ}
\end{equation}
with $J = (w^2 -\Delta^{2})/4\epsilon$, $\Delta_{a} = (w^2 +\Delta^{2})/(w^2 -\Delta^{2})$ and, $\hat{\tau}_n^x = \hat{\tau}_n^+ + \hat{\tau}_n^-$, $\hat{\tau}_n^y = -i(\hat{\tau}_n^+ -\hat{\tau}_n^-)$, $\hat{\tau}_n^z = [\hat{\tau}_n^+, \hat{\tau}_n^-]$ where $\hat{\tau}_n^\pm$ are are defined as 
  \begin{equation} 
  	\hat{\tau}_n^+ \equiv i\mathcal{G}_{2n-1}\mathcal{G}_{2n} , \quad \hat{\tau}_n^- \equiv i\mathcal{G}^\dagger_{2n-1}\mathcal{G}^\dagger_{2n}. 
  \end{equation} 
  These operators create or destroy pairs of Majorana operators on the rungs of the ladder in Fig. \ref{fig:ladder}. It can be verified (using the fermionic anti-commutation relations for $\mathcal{G}_n$ and $\mathcal{G}_n^\dagger$) that they obey the spin SU(2) commutation relations $[\hat{\tau}_n^z, \tau_{n'}^\pm ] = \pm 2 \delta_{n,n'}\tau_n^\pm$, etc., and can therefore be interpreted as Pauli matrices in the space of superoperators. 

The effective Liouvillian in Eq.~\ref{eq:effective_XXZ} is identical to the Hamiltonian for an XXZ chain, with a shift in energy so that its eigenvalues are always negative semidefinite.  Its maximum energy states are therefore the zero-energy eigenstates $\Ket{\hat{\Gamma}^{\vec{\nu} = \vec{0}}}$ and $\Ket{\hat{\Gamma}^{\vec{\nu} = \vec{1}}}$ which are related to the maximally mixed state $\Ket{\hat{\rho}_\infty} = \Ket{\hat{\Gamma}^{\vec{\nu} = \vec{0}}} / 2^N = \Ket{\hat{\mathbb{I}}}/2^N$ and the spin parity operator $\Ket{\hat{\Gamma}^{\vec{\nu} = \vec{1}}} = (-i)^N\Ket{\hat{\sigma}_z^{\otimes N}}$, respectively.

The excited energy eigenstates of the XXZ chain can be solved by employing Bethe ansatz techniques~\cite{BetheXXZ}. Finding the eigenvalue of $\mathcal{L}^{(2)}$ with smallest non-zero absolute value will therefore provide us with an estimate of the Liouvillian gap in the weak quantum limit. The closest-to-zero energy eigenstates live in the sector with one $\tau$-spin excitation, spanned by the states $\hat{\tau}_n^+ \Ket{\hat{\Gamma}^{\vec{\nu} = \vec{0}}}$ (i.e., in the $|\vec{\nu}|=2$ block, in the language of Sec. \ref{sec:block_structure}). These excitations correspond to single magnon states which have energy~\cite{Koma1997, Franchini2017}
\begin{equation}
	\mathcal{E}_{k} = \frac{w^2-\Delta^2}{\epsilon} \cos(k)-\frac{w^2+\Delta^2}{\epsilon},
\end{equation}
resulting in a relaxation gap of $\mathcal{E}_{gap} \approx -2 \Delta^2/\epsilon$ at $k=0$. We consider only the ferromagnetic case here to find the gap. The signs of $J$ and $\Delta_{a}$ depend on whether $w>\Delta$ or $w<\Delta$, producing positive or negative couplings respectively. We always have $J$ and $\Delta_{a}$ of the same sign. Moreover, the eigensystem of the model is identical under exchange of $w$ and $\Delta$ so we can obtain the anti-ferromagnetic case from the ferromagnetic one. The gap is robust as the system length $N \rightarrow \infty$ and, as the system maps directly to the XXZ chain, we can also work out the higher excitation sectors via Bethe ansatz methods. 

Recall that the parameter $\mu$, corresponding to the chemical potential of the Kitaev model, determines the phase transition. In the case of $\mu \gg 2w$, Majorana in the ladder picture (Fig.~\ref{fig:ladder}) are dimerized. Hence, there is no non-local degree of freedom. Conversely, for $\mu < 2w$, the Majorana are paired with Majorana on adjacent rungs leaving the first and last Majorana unpaired at the boundaries. This is the non-local fermion degree of freedom manifest. Moreover, in this regime the model exhibits a non-zero winding number~\cite{Kitaev2001}. 

Interestingly the $\mu$ parameter does not appear at all on the second order, contributing only at higher orders ($4^{\text{th}}$ and above).  This means that signatures of the quantum phase transition are essentially washed out in this limit.  We will show that this situation changes dramatically in the weak dissipative limit, and that the underlying effects of topology are clearly visible in the expression of the Liouvillian gap. 

\subsection{Weak dissipative limit - projecting to the kernel of $\mathbb{H}$}

We need to take a different strategy to approach the weak dissipative limit ($ \epsilon \ll w, |\Delta|,|\mu|$). To do this, we shall first define, using the free-modes of the Hamiltonian, a set of vectorized operators that lie in the kernel of $\mathcal{H}$. Then, we expand the dephasing term in this basis and show that, for low quasi-particle excitation numbers, this effective Lindbladian permits a direct solution.

Using the decomposition of quadratic Hamiltonians into normal modes as in Eq.~\ref{eq:bogoliubov}, we propose a convenient set of superoperators
\begin{equation} 
	\mathcal{B}_{n} = \sum_{j=1}^{2N} W_{nj}^* \left(\mathcal{G}_{j} + \mathcal{G}_{j}^{\dagger} \right).\label{eq:calBNM}
\end{equation} 
Acting on the identity these superoperators create the normal mode operators: 
\begin{eqnarray} 
	\Ket{\beta_n} =  \mathcal{B}_{n}  \Ket{\mathbb{I}}, \quad \Ket{\beta_n^\dagger} =  \mathcal{B}^\dagger_{n}  \Ket{\mathbb{I}} . 
\end{eqnarray} 
It is clear, therefore, that these vectorized operators $\Ket{\beta_n^\dagger \beta_{n'}} = \mathcal{B}^\dagger_{n} \mathcal{B}_{n'} \Ket{\mathbb{I}}$ are eigenstates of the Hamiltonian commutator superoperator 
\begin{equation} 
	\mathcal{H} \Ket{\beta_n^\dagger \beta_{n'}} = (\omega_n - \omega_{n'}) \Ket{\beta_n^\dagger \beta_{n'}} , 
\end{equation} 
and are supported in $\mathcal{H}_{2}$, i.e., the $|\vec{\nu}|^2 = 2$ block of $\mathcal{H}$.

\begin{figure*}[htb!]
\includegraphics[width = 2\columnwidth, page = 2]{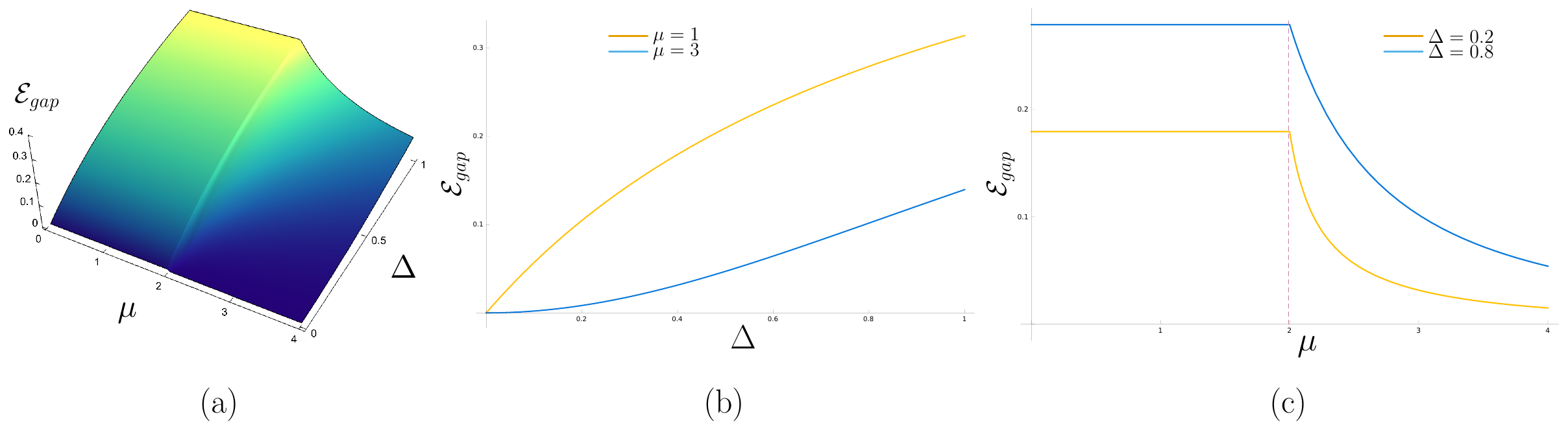}
\caption{\label{fig:GapFunction} (Color Online) (a) The Liouvillian gap for $\mu \in \lbrace 0,4\rbrace$ and $\Delta \in \lbrace 0,1\rbrace$. (b) Plot of the Liouvillian gap with respect to $\Delta$ in the topological ($\mu = 1$) and trivial ($\mu = 3$) phases respectively. (c) The gap plotted against $\mu$ at fixed $\Delta$. Here we see the change in behaviour at the phase transition $\mu = 2w$, marked by the dashed line. For all plots coherent hopping $w = 1$.
}
\end{figure*}

It was shown in \cite{Kells2015b} that to enumerate states in the kernel of $\mathcal{H}$ one can symmetrically ``super-create'' terms that have creation and annihilation operators of the same free-fermion modes $\beta$, e.g.,
\begin{equation}
	\mathcal{K}_{n}  \equiv    \mathcal{B}_{n} \mathcal{B}^{\dagger}_{n}  - \mathcal{B}^{\dagger}_{n}  \mathcal{B}_{n}.
\end{equation}
Operating with a $\mathcal{K}$ operator on the identity element (maximally mixed state) gives
\begin{equation}
	\mathcal{K}_{n} \Ket{\mathbb{I}} = \Ket{K_n} \equiv \Ket{\beta^{\phd}_n   \beta_n^\dagger - \beta_n^\dagger \beta^{\phd}_n },
\end{equation}
and from here one can span the full kernel of $\mathcal{H}$ with the states
\begin{equation}
		\Ket{K_n},\; \Ket{K_{n_{1}}, K_{n_{2}}},\; \dots, \; \Ket{K_{n_{1}},K_{n_{2}} ... , K_{n_{N}}}.
\end{equation}
Next we expand the dissipative terms of $\mathcal{L}$ in this basis and use it to gain insight into the behaviour of the Liouvillian gap in the weak dissipative limit.  Of particular interest is the behaviour of the $|\vec{\nu}|=2$ block
\begin{equation}
	[\mathcal{L}_2]_{n,m} \equiv \Bra{K_{n}}\mathcal{L}\Ket{K_{m}}.
\end{equation}
To work out the functional form for these states we  start by assuming a system with periodic boundary conditions, such that, in the Majorana basis, the normal modes (Eq.~\ref{eq:calBNM}) can be expressed using (see App.~\ref{app:BogTrans})
\begin{align}
	W_{a,k} &= \frac{-e^{ i k \lceil a/2\rceil}i^{\text{mod}(a,2)}  }{ \sqrt{2 N}} (  v_{k} + (-1)^{\text{mod}(a,2)} u_{k}),
\end{align}
where
\begin{equation}
	u_{k} = \sqrt{\frac{1}{2}+\frac{\varepsilon_k}{2 E_{k}}}, \quad
	v_{k} = -\sqrt{\frac{1}{2}-\frac{\varepsilon_k}{2 E_{k}}} e^{i \text{arg} \Delta_k },
\end{equation} 
for $\varepsilon_{k} = -\mu -2 w \cos k $, $ E_k =\sqrt{ \varepsilon_k^2 + |\Delta_k|^2},$
and $ \Delta_k = 2 i  \Delta \sin k$. For two particles this gives the state,
\begin{equation}\label{eq:two_ptcl_state}
	\Ket{K_k} = \sum_{a_1,a_2}  W_{a_1,a_2,k}   \Ket{\gamma_{a_1} , \gamma_{a_2}}
\end{equation}
with $W_{a_{1},a_{2},k} = i\text{Im}(W_{a_{1},k}W^{*}_{a_{2},k})$
\begin{equation}
	W_{a_{1},a_{2},k} = 
	\begin{cases}
		\frac{i}{N}\sin(k \tilde{\delta}_{a}),\quad \text{mod}((a_{1}-a_{2}),2) = 0,\\
		\frac{i}{N E_{k}}\left[(-1)^{\text{mod}(a_{1},2)}\varepsilon_{k}\cos(k \tilde{\delta}_{a}) + \abs{\Delta_{k}}\sin(k\tilde{\delta}_{a})\right],
	\end{cases}
\end{equation}
and $$ \tilde{\delta}_{a} \equiv \lceil\frac{a_{1}-a_{2}}{2}\rceil. $$  This expression is purely imaginary if we take $\Delta \in \mathbb{R}$.
To probe the weak dissipative limit we can now project to the 2-particle block to produce the $\mathcal{L}_{2}$  using the states $\Ket{K_{k}}$  in \eqref{eq:two_ptcl_state}. Combining the expressions, see App.~\ref{app:proj_calc}, for the states $\Ket{K_{k}}$ and the dissipator expressed in Majorana superoperators yields
\begin{equation}
	\begin{split}
		[\mathcal{L}_{2}]_{k,k^{\prime}} &= \Bra{K_{k}}\mathcal{D}\Ket{K_{k^{\prime}}},\\
	&= \epsilon\sum^{N}_{m = 1}W^{*}_{2m-1,2m,k}W_{2m-1,2m,k^{\prime}} - \epsilon N \delta_{k,k^{\prime}},\\
	&= \epsilon \sum^{N}_{m = 1} \frac{1}{N^{2}}\frac{\varepsilon_{k}\varepsilon_{k^{\prime}}}{E_{k}E_{k^{\prime}}} - \epsilon N \delta_{k,k^{\prime}},\\
	[\mathcal{L}_{2}]_{k,k^{\prime}}
	&= \frac{\epsilon}{N}\left( \tilde{\ket{\psi_{k}}}\tilde{\bra{\psi_{k^{\prime}}}} -N\right),
	\end{split}
\end{equation}
where we have defined the vector $\tilde{\ket{\psi_{k}}} \equiv \varepsilon_{k}/E_{k}$ with normalised form $\ket{\psi} = J_{\psi}^{-1}\tilde{\ket{\psi}}$ and $J_{\psi}^{2} = \braket{\tilde{\psi}}{\tilde{\psi}} = \sum_k \varepsilon_k^2/E_k^2$. In total, then, by projecting to the 2-excitation kernel $\Ket{K_{k}}$ we obtain the sub-matrix
\begin{equation}
	\mathcal{L}_{2} =  \frac{\epsilon}{N}  \left(  J_{\psi}^{2} \ket{\psi}\bra{\psi}   - N   \right),
	\label{eq:DP}
\end{equation}
from which we can directly read off the principal eigenvalue as,
\begin{equation}
	\mathcal{E}_{gap} = -\frac{\epsilon}{N} \sum_{k}  \frac{|\Delta_{k}|^2}{E_k^2} \xrightarrow{N\rightarrow \infty} -\epsilon \int dk  \frac{|\Delta_{k}|^2}{E_k^2}.
	\label{eq:DP_project}
\end{equation}
This is our approximation for the Liouvillian gap in the weak dissipative limit. Substituting $\Delta_k = 2i\Delta\sin k$, $E_k=\sqrt{\varepsilon_k^2 + \Delta_k^2}$ and $\varepsilon_k = -\mu-2w\cos k$ gives
\begin{align}\label{eq:Gap_Integral_full}
	&\mathcal{E}_{gap}=  - \epsilon \int^{\pi}_{-\pi} dk  \frac{|2 \Delta \sin (k)|^2} {  (\mu  + 2 w\cos (k))^{2} +  |2 \Delta \sin (k)|^{2}}.
\end{align}
In Appendix~\ref{app:gap_calc}, we show how this expression can be evaluated as a contour integral.  Remarkably we find that within the topological region ($|\mu| < 2w$),  the integral is independent of $\mu$ giving
\begin{equation}
	\mathcal{E}_{gap} =4 \pi  \epsilon \frac{\Delta}{w+\Delta}. 
\end{equation}
In the non-topological region  ($|\mu| >2w)$ this transitions to
\begin{equation}
	\mathcal{E}_{gap} =4 \pi \epsilon \frac{  \Delta^2} {w^2-\Delta^2} \left( \frac{|\mu|}{\sqrt{\mu^2 - 4w^2 +4 \Delta^2}} -1\right),
\end{equation}
which decays as $1/\sqrt{|\mu|}$ when $|\mu| \gg 2w$, see Figure~\ref{fig:GapFunction}.

The constant (with respect to $\mu$) relaxation gap is highly unusual and reminiscent of the behaviour of the topological index or winding number for the XY system, see e.g. \cite{StanescuBook}.  Translating into the spin language it implies that across the full ferromagnetic region the first-order response of the system is entirely independent of any applied transverse field. Conversely, when the parameters of the Hamiltonian enter the paramagnetic regime, the relaxation gap experiences a sharp drop-off as the field amplitude is made larger. Crucially, neither expression depends on the system length $N$  which therefore means that it is possible to engineer a robust and precisely controlled excitation gap in the thermodynamic limit. Such systems, with constant (in length) relaxation gaps, are often referred to as rapidly mixing~\cite{Poulin2010, Nachtergaele2011, Kastoryano2013, Lucia2015, Cubitt2015, Znidaric2015}.

In the model considered the jump operators $\ell$ are Hermitian and result in steady states $\hat{\rho}_\infty = (\hat{\Gamma}^{\vec{0}} \pm \hat{\Gamma}^{\vec{1}})/2^N$ with well-defined parity. We note that for non-Hermitian jump operators (e.g. the TASEP model of the appendix) the steady state can be quite different, and can be understood on a perturbative level as an iteratively dressed thermal state \cite{Kavanagh2022}.

We also note that some recent work has shown that the order of limits -- i.e., whether the weak dissipative limit or the thermodynamic limit is taken first -- can have a drastic effect on the Liouvillian gap. In particular, Ref. \cite{mori2023} showed that, surprisingly, if the thermodynamic limit $N\to\infty$ is taken first, then the Liouvillian gap can remain open in the weak dissipative limit $\epsilon \to 0$. However, in this section, since we take the weak dissipative limit first, the gap (in Eq. \ref{eq:Gap_Integral_full}) is proportional to $\epsilon$ in the thermodyamic limit.

\section*{Conclusion}
In this paper, we have demonstrated how operator quantization can be used to estimate the Liouvillian gap in open quantum systems - in the two extreme ratios of stochasticity (classical) and coherence (quantum). The first key idea is to transform to the canonical Majorana basis for the superoperator of the Liouvillian and exploit the block diagonal structure (associated with excitation number symmetry) therein.  In the case of Hermitian jump operators, where the steady state is the infinite temperature maximally mixed state, it is possible to solve within each block separately, without worrying about coupling to others.  Focusing primarily on the 2-excitation block of the Liouvillian, in which we see generally find the smallest gap, we are able to treat both strong classical (weak quantum ) and weak classical (weak dissipative) regimes using degenerate perturbation theory. 

In the strong classical regime, where jump operators dominate, the perturbative analysis maps to an XXZ model on the second order. Crucially there is no dependence on the external magnetic field (chemical potential in the fermionic picture) until the fourth order implying that details of the underlying Hamiltonian are essentially irrelevant at this scale. A more technical analysis that underlying Hamiltonian can however have a dramatic effect on the behaviour of the system in the limit of weak dissipative effects shows. Moreover, we show that the topological fingerprints within the XY Hamiltonian distinctly affect the behaviour of the Liovillian gap. 

Finally, it is also worth commenting on the integral expression for the complex gap function, which arises from the momentum-space parametrization of the free fermion eigenmodes of the single-particle Hamiltonian. In this latter sense, there is a clear analogy to be made with expressions that arise in the context of topological winding numbers, see e.g. \cite{StanescuBook}, whereby the topological character of the bulk physics can be condensed to a single index that also involves an integration over the full single-particle Brillouin zone.   In that situation, although a ground-state gap is necessary for the integral to be well-posed, the value of this gap is typically determined by the smallest bulk single-particle excitation energy. On the other hand, a non-zero topological index is a property of the full band structure and typically manifests in peripheral ways such as ground-state degeneracies, or the number of single-particle edge states.  One of the general insights that we now have from this analysis here is that we can see how it is {\it also} this full set of free-fermion modes that collectively determine the Liouvillian gap (and hence the steady-state relaxation times) when the classical noise is not too strong.

Our findings reveal two promising applications. First, they offer a novel method for detecting topological transitions: in the weak dissipation regime, topological phases exhibit significantly larger Liouvillian gaps compared to their non-topological counterparts at equivalent anisotropy levels. This marked difference in relaxation rates provides a distinctive experimental signature of the transition. Second, they suggest a pathway to accelerated state preparation. While the Liouvillian gap typically diminishes with increasing system size, we discovered that introducing even minimal spin anisotropy can dramatically accelerate the `cooling' process toward the desired steady state.

\begin{acknowledgments}
G.K., S.D., and K.K. acknowledge Science Foundation Ireland (SFI) for financial support through Career Development Award 15/CDA/3240. K.K. also acknowledges support from Slovenian Research Agency Grant No. J1-4385. S.D. also acknowledges support through the SFI-IRC Pathway Grant 22/PATH-S/10812. J.K.S. was supported through SFI Principal Investigator Award 16/IA/4524.
\end{acknowledgments}

\bibliographystyle{apsrev4-1}
\bibliography{Top_Fing_Gap.bib}

\onecolumngrid
\appendix

\newpage
\section{Derivation of the effective XXZ model for $\mathcal{L}^{(2)}$ in the weak quantum limit}\label{app:eff_XXZ_deriv}

Here we derive Eq.~\ref{eq:effective_XXZ}, the expression for $\mathcal{L}^{(2)}$ with the matrix elements $[\mathcal{L}^{(2)}]^{\vec{\nu}', \vec{\nu}} = \Bra{\Gamma^{\vec{\nu}'}} \mathcal{H}^2 \Ket{\hat{\Gamma}^{\vec{\nu}}}$ restricted to the kernel of $\mathcal{D}$. The kernel of $\mathcal{D}$ is spanned by the operators $\Ket{\hat{\Gamma}^{\vec{\nu}}}$ with the property that $\nu_{2n-1} = \nu_{2n}$ for all $n = 1,2,\hdots,N$.

First, we use the expression for $\mathcal{H}$ Eq.~\ref{eq:mathcal_H}, as well as the property that $\nu_{2n-1} = \nu_{2n}$ for all $n = 1,2,\hdots,N$, to compute: 
\begin{equation} 
	\mathcal{H} \Ket{\hat{\Gamma}^{\vec{\nu}}} = (\Delta + w) \sum_{n=1}^N (\mathcal{G}_{2n-1} \mathcal{G}_{2n+2}^\dagger + \mathcal{G}_{2n-1}^\dagger \mathcal{G}_{2n+2} ) \Ket{\hat{\Gamma}^{\vec{\nu}}} + (\Delta - w) \sum_{n=1}^N (\mathcal{G}_{2n} \mathcal{G}_{2n+1}^\dagger + \mathcal{G}_{2n}^\dagger \mathcal{G}_{2n+1} ) \Ket{\hat{\Gamma}^{\vec{\nu}}} . \label{eq:first_order} 
\end{equation} 
From this expression we can immediately see that the first order correction vanishes: 
\begin{equation} 
	[\mathcal{L}^{(1)}]^{\vec{\nu}', \vec{\nu}} = \Bra{\Gamma^{\vec{\nu}'}} \mathcal{H} \Ket{\hat{\Gamma}^{\vec{\nu}}} = 0 , 
\end{equation} 
when $\nu_{2n-1}' = \nu_{2n}'$, since each individual term vanishes, e.g., $\Bra{\Gamma^{\vec{\nu}'}} \mathcal{G}_{2n-1} \mathcal{G}_{2n+2}^\dagger \Ket{\Gamma^{\vec{\nu}}} = 0$.

Next, we apply the $\mathcal{H}$ superoperator to the expression in Eq.~\ref{eq:first_order} to obtain: 
\begin{eqnarray} 
	[\mathcal{L}^{(2)}]^{\vec{\nu}', \vec{\nu}} 
	&=& \Bra{\Gamma^{\vec{\nu}'}} \mathcal{H}^2 \Ket{\hat{\Gamma}^{\vec{\nu}}} \nonumber\\ 
	&=& (\Delta + w)^2 \sum_{n,n'} \Bra{\Gamma^{\vec{\nu}'}} (\mathcal{G}_{2n-1} \mathcal{G}_{2n+2}^\dagger + \mathcal{G}_{2n-1}^\dagger \mathcal{G}_{2n+2} )  (\mathcal{G}_{2n'-1} \mathcal{G}_{2n'+2}^\dagger + \mathcal{G}_{2n'-1}^\dagger \mathcal{G}_{2n'+2} ) \Ket{\hat{\Gamma}^{\vec{\nu}}} \nonumber\\ 
	&& + (\Delta - w)^2 \sum_{n,n'} \Bra{\Gamma^{\vec{\nu}'}} (\mathcal{G}_{2n} \mathcal{G}_{2n+1}^\dagger + \mathcal{G}_{2n}^\dagger \mathcal{G}_{2n+1} ) (\mathcal{G}_{2n'} \mathcal{G}_{2n'+1}^\dagger + \mathcal{G}_{2n'}^\dagger \mathcal{G}_{2n'+1} ) \Ket{\hat{\Gamma}^{\vec{\nu}}} \nonumber\\ 
	&& + (\Delta^2 - w^2) \sum_{n,n'} \Bra{\Gamma^{\vec{\nu}'}} (\mathcal{G}_{2n-1} \mathcal{G}_{2n+2}^\dagger + \mathcal{G}_{2n-1}^\dagger \mathcal{G}_{2n+2} )  ( \mathcal{G}_{2n'} \mathcal{G}_{2n'+1}^\dagger + \mathcal{G}_{2n'}^\dagger \mathcal{G}_{2n'+1} ) \Ket{\hat{\Gamma}^{\vec{\nu}}} \nonumber\\ 
	&& + (\Delta^2 - w^2) \sum_{n,n'} \Bra{\Gamma^{\vec{\nu}'}} (\mathcal{G}_{2n} \mathcal{G}_{2n+1}^\dagger + \mathcal{G}_{2n}^\dagger \mathcal{G}_{2n+1} )  ( \mathcal{G}_{2n'-1} \mathcal{G}_{2n'+2}^\dagger + \mathcal{G}_{2n'-1}^\dagger \mathcal{G}_{2n'+2} ) \Ket{\hat{\Gamma}^{\vec{\nu}}} . 
	\end{eqnarray} 
Now, thinking in terms of Fig.~\ref{fig:ladder} in the main text, only terms that involve a Majorana hopping from one rung to the next, followed by the reverse hop, will survive the projection back into the kernel. This gives: 
\begin{eqnarray} 
	[\mathcal{L}^{(2)}]^{\vec{\nu}', \vec{\nu}} 
	&=& \Bra{\Gamma^{\vec{\nu}'}} \mathcal{H}^2 \Ket{\hat{\Gamma}^{\vec{\nu}}} \nonumber\\ 
	&=& - (\Delta + w)^2 \sum_{n} \Bra{\Gamma^{\vec{\nu}'}} (\mathcal{G}_{2n-1} \mathcal{G}_{2n-1}^\dagger  \mathcal{G}_{2n+2}^\dagger \mathcal{G}_{2n+2} +  \mathcal{G}_{2n-1}^\dagger \mathcal{G}_{2n-1} \mathcal{G}_{2n+2} \mathcal{G}_{2n+2}^\dagger ) \Ket{\hat{\Gamma}^{\vec{\nu}}} \nonumber\\ 
	&& - (\Delta - w)^2 \sum_{n} \Bra{\Gamma^{\vec{\nu}'}} (\mathcal{G}_{2n} \mathcal{G}_{2n}^\dagger \mathcal{G}_{2n+1}^\dagger \mathcal{G}_{2n+1} + \mathcal{G}_{2n}^\dagger \mathcal{G}_{2n} \mathcal{G}_{2n+1} \mathcal{G}_{2n+1}^\dagger ) \Ket{\hat{\Gamma}^{\vec{\nu}}} \nonumber\\ 
	&& + 2(\Delta^2 - w^2) \sum_{n} \Bra{\Gamma^{\vec{\nu}'}} (\mathcal{G}_{2n-1} \mathcal{G}_{2n}  \mathcal{G}_{2n+1}^\dagger \mathcal{G}_{2n+2}^\dagger + \mathcal{G}_{2n-1}^\dagger \mathcal{G}_{2n}^\dagger \mathcal{G}_{2n+1} \mathcal{G}_{2n+2} ) \Ket{\hat{\Gamma}^{\vec{\nu}}} . \label{eq:second_iteration_L2} 
\end{eqnarray} 
In the last line we substitute $\hat{\tau}_n^- = i \mathcal{G}_{2n-1} \mathcal{G}_{2n}$ and $\hat{\tau}_n^+ = i \mathcal{G}_{2n-1}^\dagger \mathcal{G}_{2n}^\dagger$ to obtain:

\begin{eqnarray} 
	[\mathcal{L}^{(2)}]^{\vec{\nu}', \vec{\nu}} 
	&=& \Bra{\Gamma^{\vec{\nu}'}} \mathcal{H}^2 \Ket{\hat{\Gamma}^{\vec{\nu}}} \nonumber\\ 
	&=& - (\Delta + w)^2 \sum_{n} \Bra{\Gamma^{\vec{\nu}'}} (\mathcal{G}_{2n-1} \mathcal{G}_{2n-1}^\dagger  \mathcal{G}_{2n+2}^\dagger \mathcal{G}_{2n+2} +  \mathcal{G}_{2n-1}^\dagger \mathcal{G}_{2n-1} \mathcal{G}_{2n+2} \mathcal{G}_{2n+2}^\dagger ) \Ket{\hat{\Gamma}^{\vec{\nu}}} \nonumber\\ 
	&& - (\Delta - w)^2 \sum_{n} \Bra{\Gamma^{\vec{\nu}'}} (\mathcal{G}_{2n} \mathcal{G}_{2n}^\dagger \mathcal{G}_{2n+1}^\dagger \mathcal{G}_{2n+1} + \mathcal{G}_{2n}^\dagger \mathcal{G}_{2n} \mathcal{G}_{2n+1} \mathcal{G}_{2n+1}^\dagger ) \Ket{\hat{\Gamma}^{\vec{\nu}}} \nonumber\\ 
	&& + 2(\Delta^2 - w^2) \sum_{n} \Bra{\Gamma^{\vec{\nu}'}} ( \hat{\tau}_n^- \hat{\tau}_{n+1}^+ + \hat{\tau}_n^+ \hat{\tau}_{n+1}^- ) \Ket{\hat{\Gamma}^{\vec{\nu}}} . \label{eq:third_iteration_L2} 
\end{eqnarray}

Finally, since $\nu_{2n-1} = \nu_{2n}$ we have: 
\begin{eqnarray} 
	\mathcal{G}_{2n-1}\mathcal{G}_{2n-1}^\dagger \Ket{\hat{\Gamma}^{\vec{\nu}}} 
	&=& \mathcal{G}_{2n} \mathcal{G}_{2n}^\dagger \Ket{\hat{\Gamma}^{\vec{\nu}}} =  \mathcal{G}_{2n-1}\mathcal{G}_{2n-1}^\dagger \mathcal{G}_{2n} \mathcal{G}_{2n}^\dagger \Ket{\hat{\Gamma}^{\vec{\nu}}} = \hat{\tau}_n^- \hat{\tau}_n^+ \Ket{\hat{\Gamma}^{\vec{\nu}}} \label{eq:cheat_1} \\ \mathcal{G}_{2n-1}^\dagger\mathcal{G}_{2n-1} \Ket{\hat{\Gamma}^{\vec{\nu}}} 
	&=& \mathcal{G}_{2n}^\dagger \mathcal{G}_{2n} \Ket{\hat{\Gamma}^{\vec{\nu}}} =  \mathcal{G}_{2n-1}^\dagger\mathcal{G}_{2n-1} \mathcal{G}_{2n}^\dagger \mathcal{G}_{2n} \Ket{\hat{\Gamma}^{\vec{\nu}}} = \hat{\tau}_n^+ \hat{\tau}_n^- \Ket{\hat{\Gamma}^{\vec{\nu}}} . \label{eq:cheat_2} 
\end{eqnarray} 
We can substitute Eqs.~\ref{eq:cheat_1} and \ref{eq:cheat_2} into the first two lines of Eq. \ref{eq:third_iteration_L2} to obtain: 
\begin{eqnarray} 
	[\mathcal{L}^{(2)}]^{\vec{\nu}', \vec{\nu}} 
	&=& -(\Delta + w)^2 \sum_{n} \Bra{\Gamma^{\vec{\nu}'}} ( \hat{\tau}_n^- \hat{\tau}_{n}^+ \hat{\tau}_{n+1}^+ \hat{\tau}_{n+1}^- + \hat{\tau}_n^+ \hat{\tau}_{n}^- \hat{\tau}_{n+1}^- \hat{\tau}_{n+1}^+ ) \Ket{\hat{\Gamma}^{\vec{\nu}}} \nonumber\\ 
	&& - (\Delta - w)^2 \sum_{n} \Bra{\Gamma^{\vec{\nu}'}} ( \hat{\tau}_n^- \hat{\tau}_{n}^+ \hat{\tau}_{n+1}^+ \hat{\tau}_{n+1}^- + \hat{\tau}_n^+ \hat{\tau}_{n}^- \hat{\tau}_{n+1}^- \hat{\tau}_{n+1}^+ ) \Ket{\hat{\Gamma}^{\vec{\nu}}} \nonumber\\ 
	&& + 2(\Delta^2 - w^2) \sum_{n} \Bra{\Gamma^{\vec{\nu}'}} (\mathcal{G}_{2n-1} \mathcal{G}_{2n}  \mathcal{G}_{2n+1}^\dagger \mathcal{G}_{2n+2}^\dagger + \mathcal{G}_{2n-1}^\dagger \mathcal{G}_{2n}^\dagger \mathcal{G}_{2n+1} \mathcal{G}_{2n+2} ) \Ket{\hat{\Gamma}^{\vec{\nu}}} . \label{eq:fourth_iteration_L2} 
\end{eqnarray} 
Adding the first two lines together, and replacing the $\hat{\tau}^\pm$'s with $\hat{\tau}^{x/y/z}$'s gives our final expression: 
\begin{equation} 
\begin{split}
    	[\mathcal{L}^{(2)}]^{\vec{\nu}', \vec{\nu}} &= \Bra{\Gamma^{\vec{\nu}'}} \left[ -(\Delta^2 + w^2) \sum_n (\hat{\mathbb{I}} - \hat{\tau}_n^z\hat{\tau}_{n+1}^z ) - (\Delta^2 - w^2)\sum_n (\hat{\tau}_n^x\hat{\tau}_{n+1}^x + \hat{\tau}_n^y\hat{\tau}_{n+1}^y) \right]  \Ket{\Gamma^{\vec{\nu}'}},\\
    &= \Bra{\Gamma^{\vec{\nu}'}}(w^2 - \Delta^2)\left[ \frac{(w^2 + \Delta^2)}{(w^2 - \Delta^2)}\sum_n (\hat{\tau}_n^z\hat{\tau}_{n+1}^z -\hat{\mathbb{I}} ) + \sum_n (\hat{\tau}_n^x\hat{\tau}_{n+1}^x + \hat{\tau}_n^y\hat{\tau}_{n+1}^y) \right]  \Ket{\Gamma^{\vec{\nu}'}},\\
     &= \Bra{\Gamma^{\vec{\nu}'}}\,4\epsilon J\left[\sum_n (\hat{\tau}_n^x\hat{\tau}_{n+1}^x + \hat{\tau}_n^y\hat{\tau}_{n+1}^y) +\Delta_{a} (\hat{\tau}_n^z\hat{\tau}_{n+1}^z -\hat{\mathbb{I}} ) \right] \Ket{\Gamma^{\vec{\nu}'}},
\end{split}
\end{equation}
where we have defined the coupling $J = (w^{2} - \Delta^{2})/4\epsilon$ and anisotropy $\Delta_{a} = (w^2 + \Delta^2)/(w^2 - \Delta^2)$.


\section{Examples with dissipation modelled by exclusion processes}\label{app:Oexamples}
\subsection{Symmetric simple exclusion process}

The same expression also arises in the treatment of stochastic hopping.  Consider for example the Hermitian  process
\begin{equation}
	\begin{split}
		\label{eq:ESSEP}
			& \ell_{2x-1} = \sqrt{\epsilon} \sigma^{-}_{x} \sigma^{+}_{x+1} + \sqrt{\epsilon} \sigma^{+}_{x} \sigma^{-}_{x+1},  \\
			& \ell_{2x}\quad =  \sqrt{\epsilon} \sigma^{-}_{x} \sigma^{+}_{x+1} - \sqrt{\epsilon} \sigma^{+}_{x} \sigma^{-}_{x+1},
	\end{split}
\end{equation}
This choice of Hermitian jump operators has the nice property that it preserves what we call excitation number symmetry. When looking at the superoperator matrix representation of the Liouvillian ($\mathbb{L}$) this leads to a hierarchy of blocks that can be solved individually~\cite{Eisler2011}.

As  in the previous section  we project to the 2-particle kernel of the commutator 
\begin{equation}
	\begin{split}
		\tilde{\mathcal{L}}^{(2)}_{k,k'}
		&=  \Bra{K_k} \mathcal{D} \Ket{K_{k'}}, \\
		&= \epsilon \frac{4}{N} \cdot \left( \frac{\varepsilon_k \varepsilon_{k'}}{E_k E_{k'}}  +  \frac{\Delta_k \Delta_{k'} \sin{k} \sin {k'} }{E_k E_{k'}}\right).
	\end{split}
\end{equation}
from which we can extract the non-trivial eigenvalues exactly from the two-level model
\begin{equation}
	H_{\text{eff}}= \epsilon\frac{4}{N} \begin{bmatrix}
	J_{\psi}^{2}  & \braket{\psi}{\phi} \\
	\braket{\phi}{\psi} & J_{\phi}^{2}
	\end{bmatrix}  - 4 \epsilon I_2
\end{equation}
where  $\tilde{\ket{\psi}}_{k} = \varepsilon_k/E_k$ and $\tilde{\ket{\phi}}_{k}   = \sin k \times \Delta_k/E_k$ and  $J_{\psi}^{2} = \braket{\tilde{\psi}}{\tilde{\psi}} = \sum_k \varepsilon_k^2/E_k^2$ , $J_{\phi}^{2} = \braket{\tilde{\phi}}{\tilde{\phi}} = \sum_k (\sin k )^2 \Delta_k^2/E_k^2$.  A first order estimate of the gap is then
\begin{equation}
	\mathcal{E}_{gap} =  \frac{ 4 \epsilon}{N}  J_{\psi}^{2} - \epsilon = - \frac{4 \epsilon}{N}  \sum_k \frac{ |\Delta_k|^2} {E_k^2},
\end{equation}
which, as $N \rightarrow \infty$, is $4$ times the expression given in Eq.~\ref{eq:DP_project}. In the next subsection we also show how the same expression arises, again on a perturbative level, for the TASEP dissipator that encodes non-symmetric and non-Hermitian stochastic hopping.
\subsection{Totally asymmetric simple exclusion process} 

In this section we include more details on another process, namely the TASEP, that produces a similar Liouvillian gap when combined with the XY Hamiltonian. Some of the discussion is of further relevance to the related SSEP model examined in the main text.

Temme et. al.~\cite{Temme2012} considered the totally asymmetric simple exclusion process (TASEP) with coherent hopping (i.e. $\hat{H}|_{\Delta = 0}$, the $XX$ model). In the bulk the model consists of $N-1$ stochastic terms,
\begin{equation} 
	\ell_{x} = \sqrt{\gamma} \sigma^-_x \sigma^+_{x+1},  \quad \forall x \in \lbrace 1,\dots, N-1 \rbrace, 
	\label{eq:TASEPBulk}
\end{equation}
and two boundary terms which specify particle creation (hop on of spin downs) on the left hand side and particle annihilation (hopping off of spin down) on the right
\begin{equation}
	\ell_{N} = \sqrt{\alpha} \sigma_{1}^{-} , \quad  \ell_{N+1} =\sqrt{\beta} \sigma_{N}^{+}. 
	\label{eq:TASEPBoundary}
\end{equation}

\begin{figure*}[htb!]
\includegraphics[width=1\columnwidth, page = 1]{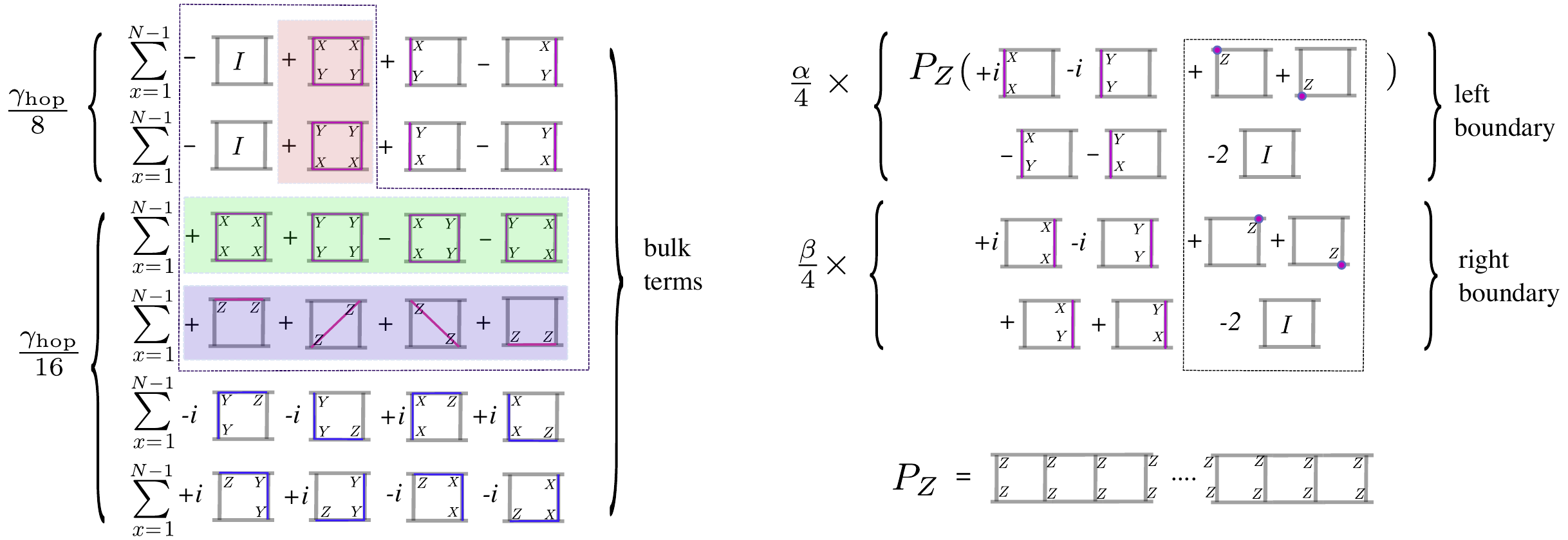}
\caption{\label{fig:Lcan} (Color Online) Full  TASEP  Linbladian superoperator in the canonical basis. We use the short hand $X=\sigma^x, Y=\sigma^y, Z=  \sigma^z $ for the Pauli operators in the canonical basis, see~\cite{Kells2015b}.   Only terms highlighted commute with the excitation number operator $\mathcal{N}$. For the Hermitian SSEP  only these highlighted terms appear - albeit multiplied by an overall constant of $4$.}
\end{figure*}

The full Lindbladian operator for the TASEP is given diagrammatically in Figure~\ref{fig:Lcan}. For TASEP we typically assume open boundary conditions for both Hamiltonian and stochastic terms. To produce the precise effective description we could then use single particle fermionic operators from this open boundary scenario. However, to simplify things we instead assume that we can use  periodic momentum fermionic creation and annihilation operators to make the zero excitation energy kernel of $\mathcal{H}$. The situation is more complicated here since the superoperator matrix is not block diagonal but using the arguments of~\cite{Kavanagh2022} we note that we can project onto the $s=2$ block for small $\gamma, \alpha, \beta$ and end up with a situation very similar to 
Eq.~\ref{eq:ESSEP}
\begin{equation}
\tilde{\mathcal{L}}^{(2)} = \gamma\frac{N-1}{N^2} \times \left(  J_\psi \ket{\psi}\bra{\psi}   + J_\phi \ket{\phi} \bra{\phi}   \right) -  (\gamma -\frac{\gamma}{N} +\frac{\alpha+\beta}{N}),
	\label{eq:TASEP_project}
\end{equation}
which, using essentially the same analysis in the main text, leads to  a gap that tends to $\mathcal{E}_{gap}$ as defined in the main text in the $N  \rightarrow \infty$ limit.

\section{The four excitation number gap}

\begin{figure*}[htb!]
\includegraphics[width=1\columnwidth, page = 3]{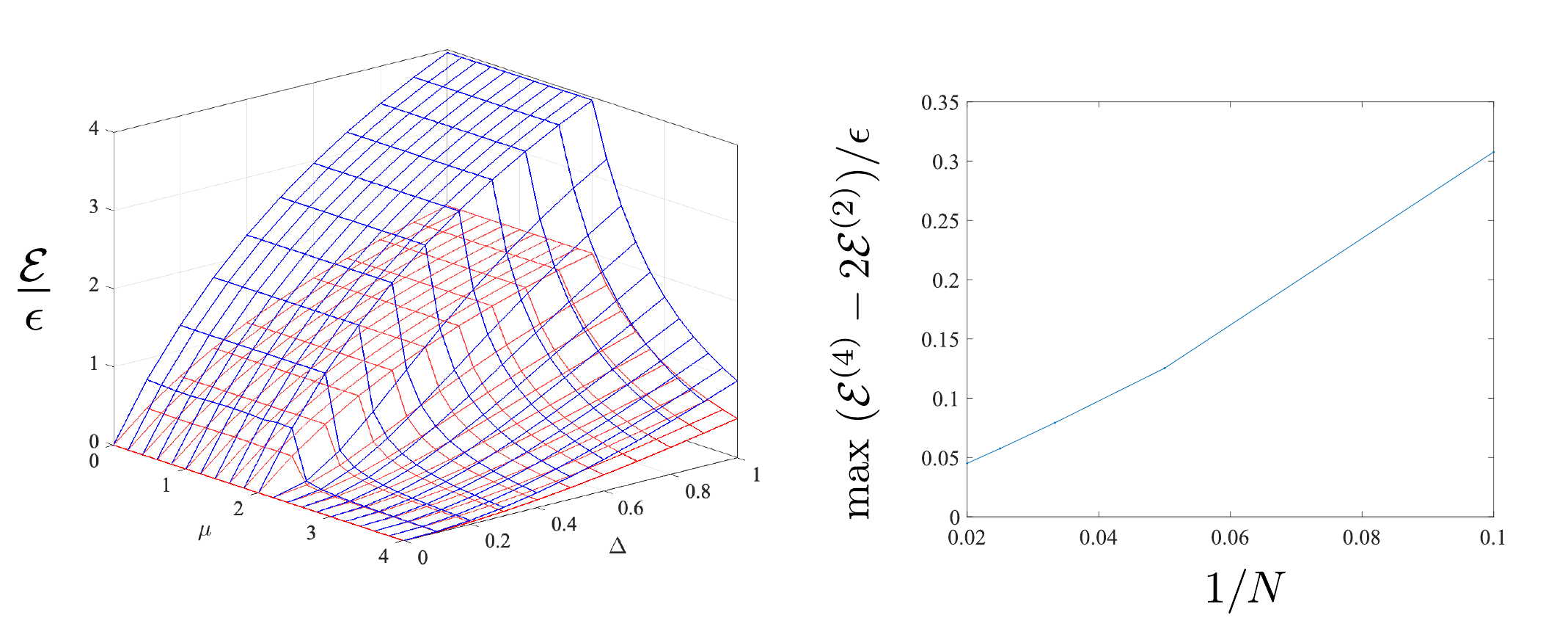}
\caption{\label{fig:G4G2} (Color Online) Smallest relaxation gaps of $|\vec{\nu}|^2 =2$ (red) and$|\vec{\nu}|^2 =4$ (blue) sectors for onsite bulk dephasing in the weak $\epsilon \rightarrow 0$ classical limit. Left: The parameterized gaps overlaid for $N=50$. Right: The $|\vec{\nu}|^2 =4$ becomes approximately equal to twice the $|\vec{\nu}|^2 =2$ gap as $N$ is increased.}
\end{figure*}

A key assumption in the main text is that the $|\vec{\nu}|^2 =2$ sector generically contains the smallest gap. We do not have an analytical proof for this. However we can analyse here how the  $|\vec{\nu}|^2 =4$ gap scales, for simplicity restricting the discussion to just the case of dephasing.   The general result is that, although we have not been able to find a simple representation in terms of a small number of projections for the $|\vec{\nu}|^2 =4$ case, the lowest energy gap nonetheless approaches {\it twice} the  $|\vec{\nu}|^2 =2$ gap as the system size is increased, see Figure \ref{fig:G4G2}. A similar analysis suggests that the pattern holds for higher sectors, underlining our justification for focusing on the $|\vec{\nu}|^2 =2$ sector.

\section{Modified Bogoliubov Transformation}\label{app:BogTrans}

On the $N$ rung ladder we can use a Bogoliubov transformation, with sites labelled by $r = (x,y)$, whereby $y = 0$ denotes the lower portion of the ladder, $y=1$ the upper portion and $x=1,\dots,N$ the rung. The free fermion modes are defined by
\begin{equation}
	\beta^{\dagger}_{k} =\sum_{r} W_{r,k} \gamma_{r},
\end{equation}
where
\begin{equation}
	W_{r,k} = \frac{-e^{ i k x}  }{ \sqrt{2 N_{x}}} ( i^{y} v_{k} + (-i)^{y} u_{k}).
\end{equation}
Unfurling the $N$ rung ladder to a $2N$ site chain we need to replace the indexing using $r$ with another site label denoted by $a\in\{1,\dots,2N\}$. Starting from the expressions above we can note that using the numbering convention of Fig.~\ref{fig:ladder}, for a given $r=(x,y)$ we have:
\begin{equation}
	\begin{split}
		x &= \text{ceil}\left(\frac{a}{2}\right),\\
		y &= \text{mod}(a,2).
	\end{split}
\end{equation}
Thus, the new expression for the free modes will now use
\begin{equation}
	\begin{split}
			W_{a,k} &= \frac{-e^{ i k \left\lceil a/2\right\rceil}  }{ \sqrt{2N}} ( i^{\text{mod}(a,2)} v_{k} + (-i)^{\text{mod}(a,2)} u_{k}),\\
			&= \frac{-e^{ i k \left\lceil a/2\right\rceil} (-i)^{\text{mod}(a,2)} }{ \sqrt{2N}} ( u_{k} + (-1)^{\text{mod}(a,2)}v_{k}).
	\end{split}
\end{equation}
Further to construct symmetric two particle states such as $\Ket{K_{k}} = \Ket{\beta_{k}\beta_{k}^{\dagger}-\beta_{k}^{\dagger}\beta_{k}}$, we need to determine
\begin{equation}
\begin{split}
		\Ket{K_{k}} &= \sum_{a_{1},a_{2}}\left(W_{a_{	1},k}W^{*}_{a_{2},k}-W^{*}_{a_{1},k}W_{a_{2},k}\right)\Ket{\gamma_{a_{1}}\gamma_{a_{2}}},\\
		&= \sum_{a_{1},a_{2}}i\text{Im}\left(W_{a_{1},k}W^{*}_{a_{2},k}\right)\Ket{\gamma_{a_{1}}\gamma_{a_{2}}},
\end{split}
\end{equation}
since for some $s,t\in\mathbb{C}$, it is easy to see that $s\cdot t^{*}-s^{*}\cdot t = i\text{Im}(s\cdot t^{*})$. Then, with this in mind we can find 
\begin{align*}
	W_{a_{1},k}W^{*}_{a_{2},k} 
	&= \left[\frac{-e^{ i k \left\lceil a/2\right\rceil} (-i)^{\text{mod}(a_{1},2)} }{ \sqrt{2N}} ( u_{k} + (-1)^{\text{mod}(a_{1},2)}v_{k})\right]\left[\frac{-e^{- i k \left\lceil a/2\right\rceil} (+i)^{\text{mod}(a_{2},2)} }{ \sqrt{2N}} ( u_{k} + (-1)^{\text{mod}(a_{2},2)}v^{*}_{k}) \right],\\
	&= \frac{e^{ i k (\left\lceil a_{1}/2\right\rceil-\left\lceil a_{2}/2\right\rceil)}i^{\text{mod}(a_{1},2)+\text{mod}(a_{2},2)}}{2N}\left[ (-1)^{\text{mod}(a_{1},2)} u_{k} + v_{k}\right]\left[u_{k} +   (-1)^{\text{mod}(a_{2},2)}v^{*}_{k} \right],\\
	&= \frac{e^{ i k \tilde{\delta}_{a}}i^{\text{mod}(a_{1},2)+\text{mod}(a_{2},2)}}{2N}\left[|u|^{2}_{k}(-1)^{\text{mod}(a_{1},2)} + u_{k}v_{k} + u_{k}v^{*}_{k}(-1)^{\text{mod}(a_{1},2)+\text{mod}(a_{2},2)}+  |v|^{2}_{k}(-1)^{\text{mod}(a_{2},2)}\right],
\end{align*}
where we define $\tilde{\delta}_{a} \equiv \left\lceil a_{1}/2\right\rceil-\left\lceil a_{2}/2\right\rceil$. Now consider if $a_{1}, a_{2}$ have equal or distinct parity. If equal then $\text{mod}(a_{1},2)+\text{mod}(a_{2},2) = 2\text{ mod}(a_{1},2)$ yielding,
\begin{equation}
	W_{a_{1},k}W^{*}_{a_{2},k} = \frac{e^{ i k \tilde{\delta}_{a}}(-1)^{\text{mod}(a_{1},2)}}{2N}\left[ (-1)^{\text{mod}(a_{1},2)}(|u|^{2}_{k} + |v|^{2}_{k}) + u_{k}(v_{k} + v^{*}_{k})\right].
\end{equation}
Recall that 
\begin{equation}
	u_{k} = \sqrt{\frac{1}{2}+\frac{\varepsilon_k}{2 E_{k}}}, \quad
	v_{k} = -\sqrt{\frac{1}{2}-\frac{\varepsilon_k}{2 E_{k}}} e^{i \text{arg} \Delta_{k} },
\end{equation} 
and $\varepsilon_{k} = -\mu -2 t \cos k $ and $ E_k =\sqrt{ \varepsilon_k^2 + |\Delta_k|^2},$. Moreover, $u_{k}^{*} = u_{k}$, $|u|^{2}_{k} + |v|^{2}_{k} = 1$, $|u|^{2}_{k} - |v|^{2}_{k} = \varepsilon_{k}/E_{k}$ and $E^{2}_{k}-\varepsilon^{2}_{k} = |\Delta_{k}|^{2}$. Then we have
\begin{align}
		W_{a_{1},k}W^{*}_{a_{2},k} &= \frac{e^{ i k \tilde{\delta}_{a}}}{2N}\left[1 + (u_{k}(v_{k}+v^{*}_{k}))(-1)^{\text{mod}(a_{1},2)}\right],\\
		&= \frac{e^{ i k \tilde{\delta}_{a}}}{2N}\left[1 - \sqrt{\frac{1}{2}+\frac{\varepsilon_k}{2 E_{k}}}\sqrt{\frac{1}{2}-\frac{\varepsilon_k}{2 E_{k}}}\left(e^{i \text{arg} \Delta_k }+e^{-i \text{arg} \Delta_k }\right)(-1)^{\text{mod}(a_{1},2)} \right],\\
		&= \frac{e^{ i k \tilde{\delta}_{a}}}{2N}\left[1 - \frac{1}{E_{k}}\sqrt{E^{2} - \varepsilon^{2}_{k}}\cos\left(\text{arg} \Delta_k\right)(-1)^{\text{mod}(a_{1},2)} \right],\\
		&= \frac{e^{ i k \tilde{\delta}_{a}}}{2N}\left[1 - \frac{|\Delta_{k}|}{E_{k}}\cos\left(\text{arg} \Delta_k\right)(-1)^{\text{mod}(a_{1},2)} \right].
\end{align}
Then finally for the equal parity case we have
\begin{equation}
	\begin{split}
		i\text{Im}\left(W_{a_{1},k}W^{*}_{a_{2},k}\right) &= \frac{i}{N}\sin(k \tilde{\delta}_{a})\left[1 - \frac{|\Delta_{k}|}{E_{k}}\cos\left(\text{arg} \Delta_k\right)(-1)^{\text{mod}(a_{1},2)} \right],\\
		&= \frac{i}{N}\sin(k \tilde{\delta}_{a}), \text{ for } \Delta_{k} \in i\mathbb{R}.
	\end{split}
\end{equation}
This is always the case when $\Delta \in \mathbb{R}$ as $\Delta_{k} = 2i\Delta\sin k$. Next we consider the case where $a_{1}$ and $a_{2}$ have different parity. In this case $\text{mod}(a_{1},2)+\text{mod}(a_{2},2) = 1$, in particular $(-1)^{\text{mod}(a_{2},2)} = -(-1)^{\text{mod}(a_{1},2) }$,
\begin{align}
	W_{a_{1},k}W^{*}_{a_{2},k} &= \frac{e^{ i k \tilde{\delta}_{a}}i}{2N}\left[|u|^{2}_{k}(-1)^{\text{mod}(a_{1},2)} + u_{k}v_{k} + u_{k}v^{*}_{k}(-1)^{\text{mod}(a_{1},2)+\text{mod}(a_{2},2)}-  |v|^{2}_{k}(-1)^{\text{mod}(a_{1},2)}\right],\\
	&= \frac{ie^{ i k \tilde{\delta}_{a}}}{2N}\left[(|u|^{2}_{k}-  |v|^{2}_{k})(-1)^{\text{mod}(a_{1},2)} + u_{k}(v_{k} - v^{*}_{k})\right],\\
	&=\frac{ie^{ i k \tilde{\delta}_{a}}}{2N}\left[\frac{\varepsilon_{k}}{E_{k}}(-1)^{\text{mod}(a_{1},2)} -\sqrt{\frac{1}{2}+\frac{\varepsilon_k}{2 E_{k}}}\sqrt{\frac{1}{2}-\frac{\varepsilon_k}{2 E_{k}}}\left(e^{i \text{arg} \Delta_k }-e^{-i \text{arg} \Delta_k }\right) \right],\\
	&= \frac{ie^{ i k \tilde{\delta}_{a}}}{2N}\left[\frac{\varepsilon_{k}}{E_{k}}(-1)^{\text{mod}(a_{1},2)} -i\frac{|\Delta_{k}|}{E_{k}}\sin\text{arg} \Delta_{k}  \right],\\
	&= \frac{ e^{ i k \tilde{\delta}_{a}}}{2NE_{k}}\left[i\varepsilon_{k}(-1)^{\text{mod}(a_{1},2)} + |\Delta_{k}|\sin\text{arg} \Delta_{k}  \right],
\end{align}
Then we can obtain
\begin{equation}
	\begin{split}
		i\text{Im}\left(W_{a_{1},k}W^{*}_{a_{2},k}\right) &= \frac{i}{NE_{k}}\left[(-1)^{\text{mod}(a_{1},2)}\varepsilon_{k}\cos(k \tilde{\delta}_{a}) + |\Delta_{k}|\sin(k \tilde{\delta}_{a})\sin\text{arg} \Delta_{k}\right],\\
		&= \frac{i}{NE_{k}}\left[(-1)^{\text{mod}(a_{1},2)}\varepsilon_{k}\cos(k \tilde{\delta}_{a}) + |\Delta_{k}|\sin(k \tilde{\delta}_{a}) \right], \text{ for } \Delta_{k} \in \mathbb{C}.
	\end{split}
\end{equation}
Thus we have the result stated in the main text
\begin{equation}
	W_{a_{1},a_{2},k} = 
	\begin{cases}
		\frac{i}{N}\sin(k \tilde{\delta}_{a}),\quad \text{mod}((a_{1}-a_{2}),2) = 0,\\
		\frac{i}{N E_{k}}\left[(-1)^{\text{mod}(a_{1},2)}\varepsilon_{k}\cos(k \tilde{\delta}_{a}) + \abs{\Delta_{k}}\sin(k\tilde{\delta}_{a})\right],
	\end{cases}
\end{equation}

\section{Kernel Projection Calculation}\label{app:proj_calc}

We begin with the form of the dissipation in the matrix representation of superoperators
\begin{equation}
  \mathcal{D} = \epsilon \sum_{n = 1}^{N} (\left[\mathcal{G}_{2n-1}^{\dagger},\mathcal{G}_{2n-1}\right]\left[\mathcal{G}_{2n}^{\dagger},\mathcal{G}_{2n}\right] - \mathbb{I}),
\end{equation}
and note the definition of the free fermion modes using the same Majorana superoperators
\begin{equation}
	\mathcal{B}_{n} = \sum_{j=1}^{2N} W_{nj}^* \left(\mathcal{G}_{j}+\mathcal{G}^{\dagger}_{j}\right),
\end{equation}
Note that in the $r = (x,y)$ indexing system, those $r = (x,1)$ are odd in $n$ indexing and those $r = (x,2)$ are even in $n$ indexing. Recall that the $|\nu|^{2} = 2$ states $\mathcal{K}$ can be defined by $\mathcal{G}$ as:
\begin{equation}
	\mathcal{K}_{k}\Ket{I} \equiv \sum^{N}_{a,a^{\prime} = 1} W_{2a-1,2a^{\prime}, k}\left(\mathcal{G}_{2a-1}+\mathcal{G}^{\dagger}_{2a-1}\right)\left(\mathcal{G}_{2a^{\prime}}+\mathcal{G}^{\dagger}_{2a^{\prime}}\right)\Ket{I},
\end{equation}
where
\begin{equation}
	W_{a_{1},a_{2},k} = 
	\begin{cases}
		\frac{i}{N}\sin(k \tilde{\delta}_{a}), &\text{mod}((a_{1}-a_{2}),2) = 0,\\
		\frac{i}{N E_{k}}\left[(-1)^{\text{mod}(a_{1},2)}\varepsilon_{k}\cos(k \tilde{\delta}_{a}) + \abs{\Delta_{k}}\sin(k\tilde{\delta}_{a})\right], &\text{otherwise}.
	\end{cases},
\end{equation}
and $ \tilde{\delta}_{a} \equiv \lceil\frac{a_{1}}{2}\rceil - \lceil\frac{a_{2}}{2}\rceil. $ Then the object we need to calculate is the principal eigenvalue of $\mathcal{L}^{(2)}_{k,k^{\prime}} = \Bra{K_{k}}\mathcal{D}\Ket{K_{k^{\prime}}}$, 
\begin{equation}
	\mathcal{L}^{(2)}_{k,k^{\prime}} = \epsilon\sum^{N}_{a,a^{\prime}=1}\sum^{N}_{b,b^{\prime} = 1}\sum^{N}_{n = 1} W^{*}_{2a-1,2a^{\prime},k}W_{2b-1,2b^{\prime},k^{\prime}} \underbrace{\Bra{\gamma_{2a^{\prime}}\gamma_{2a-1}}\left[\mathcal{G}_{2n-1}^{\dagger},\mathcal{G}_{2n-1}\right]\left[\mathcal{G}_{2n}^{\dagger},\mathcal{G}_{2n}\right]\Ket{\gamma_{2b-1}\gamma_{2b^{\prime}}}}_{\text{Comm}} - \epsilon\sum_{n = 1}^{N}\BraKet{K_{k}}{K_{k^{\prime}}}.
\end{equation}
Focus first on the commutators, notice that we can write them in the following way
\begin{equation}
	\begin{split}
		 \text{Comm} &= \Bra{\gamma_{2a^{\prime}}\gamma_{2a-1}}\left[\mathcal{G}_{2n-1}^{\dagger},\mathcal{G}_{2n-1}\right]\left[\mathcal{G}_{2n}^{\dagger},\mathcal{G}_{2n}\right]\Ket{\gamma_{2b-1}\gamma_{2b^{\prime}}},\\
		  &= \Bra{\gamma_{2a^{\prime}}\gamma_{2a-1}}\left(\mathcal{G}_{2n-1} + \mathcal{G}^{\dagger}_{2n-1}\right)\left(\mathcal{G}_{2n-1} - \mathcal{G}^{\dagger}_{2n-1}\right)\left(\mathcal{G}_{2n} + \mathcal{G}^{\dagger}_{2n}\right)\left(\mathcal{G}_{2n} - \mathcal{G}^{\dagger}_{2n}\right)\Ket{\gamma_{2b-1}\gamma_{2b^{\prime}}},\\
		  &= -\Bra{\gamma_{2a^{\prime}}\gamma_{2a-1}}\left(\mathcal{G}_{2n-1} + \mathcal{G}^{\dagger}_{2n-1}\right)\left(\mathcal{G}_{2n} + \mathcal{G}^{\dagger}_{2n}\right)\left(\mathcal{G}_{2n-1} - \mathcal{G}^{\dagger}_{2n-1}\right)\left(\mathcal{G}_{2n} - \mathcal{G}^{\dagger}_{2n}\right)\Ket{\gamma_{2b-1}\gamma_{2b^{\prime}}},\\
		  &= \Bra{\gamma_{2a^{\prime}}\gamma_{2a-1}}\left(\mathcal{G}_{2n-1} + \mathcal{G}^{\dagger}_{2n-1}\right)\left(\mathcal{G}_{2n} + \mathcal{G}^{\dagger}_{2n}\right)\left(\mathcal{G}_{2n-1} - \mathcal{G}^{\dagger}_{2n-1}\right)\delta_{b^{\prime},n}\Ket{\gamma_{2b-1}},\\
		  &= \Bra{\gamma_{2a^{\prime}}\gamma_{2a-1}}\left(\mathcal{G}_{2n-1} + \mathcal{G}^{\dagger}_{2n-1}\right)\left(\mathcal{G}_{2n} + \mathcal{G}^{\dagger}_{2n}\right)\delta_{b,n}\delta_{b^{\prime},n}\Ket{\mathbb{I}},\\
		  &= \delta_{b,n}\delta_{b^{\prime},n}\BraKet{\gamma_{2a^{\prime}}\gamma_{2a-1}}{\gamma_{2n-1}\gamma_{2n}},\\
		  &= \delta_{a,n}\delta_{a^{\prime},n}\delta_{b,n}\delta_{b^{\prime},n}.
	\end{split}
\end{equation}
The second expression for Comm has additional terms compared to the first but these are zero inside the expectation between 2-excitation states. Now reinserting this into the expression for $\mathcal{L}^{(2)}_{k,k^{\prime}}$ we obtain
\begin{equation}
	\begin{split}
		\mathcal{L}^{(2)}_{k,k^{\prime}} &= \epsilon\sum^{N}_{a,a^{\prime}=1}\sum^{N}_{b,b^{\prime} = 1}\sum^{N}_{n = 1} W^{*}_{2a-1,2a^{\prime},k}W_{2b-1,2b^{\prime},k^{\prime}} \delta_{a,n}\delta_{a^{\prime},n}\delta_{b,n}\delta_{b^{\prime},n} - \epsilon\sum_{n = 1}^{N}\BraKet{K_{k}}{K_{k^{\prime}}},\\
		&= \epsilon\sum^{N}_{n = 1}\left( W^{*}_{2n-1,2n,k}W_{2n-1,2n,k^{\prime}} - \delta_{k,k^{\prime}}\right).
	\end{split}
\end{equation}
Now note the simplification of $W_{a_{1},a_{2},k}$, for $a_{1} = 2n-1$, $a_{2} = 2n$ and hence, $\tilde{\delta_{a}} = 0$
\begin{equation}
	W_{2n-1,2n,k} = \frac{-i\varepsilon_{k}}{N E_{k}}
\end{equation}
We are able now to write the	 elements of the $|\nu|^{2} = 2$ block of the Liouvillian as
\begin{equation}
	\mathcal{L}^{(2)}_{k,k^{\prime}} = \epsilon \sum^{N}_{n = 1}\left( \frac{1}{N^{2}}\frac{\varepsilon_{k}\varepsilon_{k^{\prime}}}{E_{k}E_{k^{\prime}}} - \delta_{k,k^{\prime}}\right),
\end{equation}
which in compact form as a matrix is given by
\begin{equation}
	\mathcal{L}^{(2)} = \frac{\epsilon}{N}\left( \tilde{\ket{\psi}}\tilde{\bra{\psi}} -N\right),
\end{equation}
for $\tilde{\ket{\psi}}_{k} = \varepsilon_{k}/E_{k}$. Extracting the principal eigenvalue of this matrix then leaves us with the following expression for the Liouvillian gap
\begin{equation}
	\mathcal{E}_{gap} = -\frac{\epsilon}{N} \sum_{k}  \frac{|\Delta_{k}|^2}{E_k^2}.\end{equation}

\section{Evaluation of the Gap integral}\label{app:gap_calc}

\begin{figure*}
\centering
\includegraphics[width=1\columnwidth, page = 4]{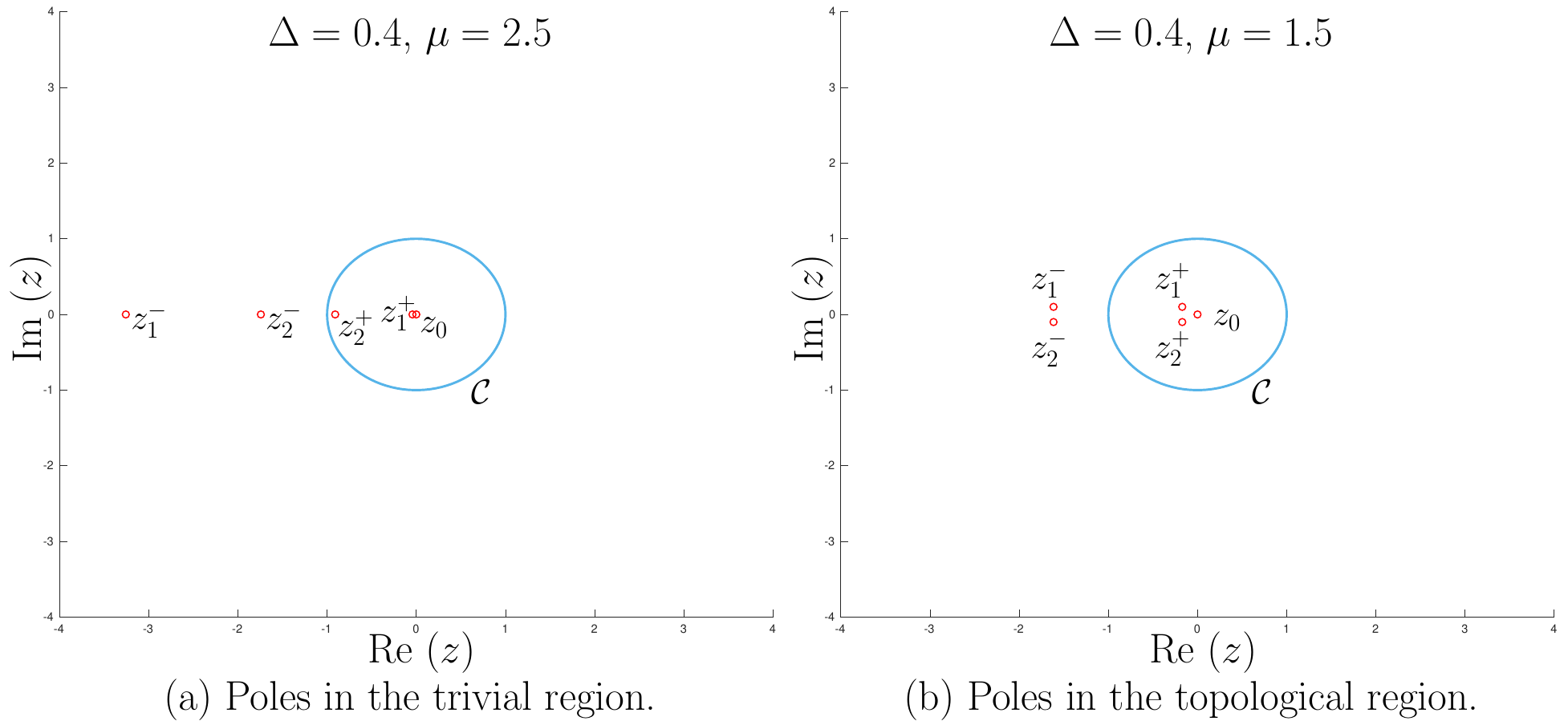}
\caption[]{\label{fig:Poles} (Color Online)  There is a subtle difference between the the pole structure of the gap integrand in the topological and non-topological regimes. In both cases the poles of relevance are those labelled as $z_{0}$, $z_{1}^{+}$ and $z_{2}^{+}$. We can see that the $+$-poles lift off the real axis and become complex when $\mu < 2w$ \textit{i.e.} inside the topological region.   }
\end{figure*}

Here we present an approach to calculate the gap integral Eq.~\ref{eq:DP_project}, by contour integral. We write it explicitly as it has appeared thus far, with minor modifications,
\begin{equation}\label{eq:Gap_integral_explicit}
	\mathcal{E}_{gap} = -\epsilon \int^{\pi}_{-\pi} dk \frac{|2 \Delta \sin (k)|^2} {  (\mu  + 2 w \cos (k))^{2} +  |2 \Delta \sin (k)|^{2}}.
\end{equation}
We begin with a change of variables, trading $k$ for $z$ in the following way:
\begin{equation}
	z = \exp(i k),\; dz = i e^{ik} dk,\; dk = -iz^{-1}dz.
\end{equation}
This then gives
\begin{equation}
	\mathcal{E}_{gap} = -i\epsilon\oint dz \frac{\Delta^{2}(2z^{2} - z^{4} -1)}{z^{3}(\mu^{2} + 2(w^{2}+\Delta^{2})) +2w\mu(z^{4}+z^{2}) +(w^{2}-\Delta^{2})(z^{5} + z) }.
\end{equation}
After some rearrangement the integral is written as
\begin{equation}
		\mathcal{E}_{gap} = -i\Delta^{2}\epsilon\oint dz \frac{(z^{2}-1)^{2}}{(w^{2}-\Delta^{2})z^{5} +2w\mu z^{4} + (\mu^{2} + 2(w^{2}+\Delta^{2}))z^{3} + 2w\mu z^{2} +(w^{2}-\Delta^{2})z }.
\end{equation}
At this point the integrand is in a suitable form to discuss which values of $z$ are poles. In particular one can see that the all come from the denominator which we find from solving the equation
\begin{equation}
	(w^{2}-\Delta^{2})z^{5} +2w\mu z^{4} + (\mu^{2} + 2(w^{2}+\Delta^{2}))z^{3} + 2w\mu z^{2} +(w^{2}-\Delta^{2})z = 0
\end{equation}
Solving this question gives, in addition to the obvious pole at zero, four others which come in two pairs. In total, the poles of the integral are:
\begin{align}
	z_{0} &= 0,\\
	z_{1}^{\pm} &= -\frac{1}{4(w^{2}-\Delta^{2})}\left(2w\mu +2\Delta\zeta\right) 
	 \pm\frac{1}{2(w^{2}-\Delta^{2})}\sqrt{\mu^{2}\Delta^{2} +w^{2}\zeta^{2} + 2w\mu\Delta\zeta } ,\\
	z_{2}^{\pm} &=  -\frac{1}{4(w^{2}-\Delta^{2})}\left(2w\mu +2\Delta\zeta\right) 
	 \pm\frac{1}{2(w^{2}-\Delta^{2})}\sqrt{\mu^{2}\Delta^{2} +w^{2}\zeta^{2} - 2w\mu\Delta\zeta } ,
\end{align}
where we have introduced another variable $\zeta$ which we define as
\begin{equation}
	\zeta \equiv \sqrt{\mu^{2}-4w^{2}+4\Delta^{2}}.
\end{equation}
Having obtained the poles explicitly we can next compute the residue of the integral at each of the poles using the $\zeta$ variable
\begin{align}\label{eq:ResiduesHamParameters}
	\text{Res}(z_{0}) &= \frac{1}{w^{2}-\Delta^{2}},\\
	\text{Res}(z_{1}^{\pm}) &= \frac{\mp 1}{2\Delta(w^{2}-\Delta^{2})\zeta}\sqrt{ w^2\zeta^{2} +\mu^{2}\Delta^{2} + 2\mu w\Delta\zeta } ,\\
	\text{Res}(z_{2}^{\pm}) &= \frac{\pm 1}{2\Delta(w^{2}-\Delta^{2})\zeta}\sqrt{ w^2\zeta^{2} +\mu^{2}\Delta^{2} - 2\mu w\Delta\zeta }.
\end{align}
We can write these residues in a final compact form leading to the expressions:
\begin{align}\label{eq:ResiduesZeta}
	\text{Res}(z_{0}) &= \frac{1}{w^{2}-\Delta^{2}},\\
	\text{Res}(z_{1}^{\pm}) &= \mp\frac{ \abs{ w\zeta + \mu\Delta }}{2\Delta(w^{2}-\Delta^{2})\zeta} ,\nonumber\\
	\text{Res}(z_{2}^{\pm}) &= \pm\frac{ \abs{ w\zeta - \mu\Delta }}{2\Delta(w^{2}-\Delta^{2})\zeta} \nonumber.
\end{align}
Finally, recall the residue theorem
\begin{equation}
	\oint_{c}f(z)dz = 2\pi i \sum_{k} \text{Res}(f,a_{k}),
\end{equation}
for $c$ a positively oriented simple closed curve (here the unit circle at the origin) and $a_{k}$ a pole in the interior of the curve. These poles are $z_{0}$, $z^{+}_{1}$ and $z^{+}_{2}$ which yield
\begin{equation}
	\begin{split}
		\mathcal{E}_{gap} 
		&= -\frac{4\pi\epsilon\Delta^{2}}{w^{2}-\Delta^{2}} - \frac{\Delta\pi \abs{ w\zeta + \mu\Delta }}{(w^{2}-\Delta^{2})\zeta} +\frac{ \Delta\pi\abs{ w\zeta - \mu\Delta }}{(w^{2}-\Delta^{2})\zeta},\\
		&= \frac{4\pi\epsilon\Delta^{2}}{w^{2}-\Delta^{2}}\left(-1 + \frac{1}{2\Delta\zeta}\left(\abs{ w\zeta + \mu\Delta } -\abs{ w\zeta - \mu\Delta }\right) \right)
	\end{split}
\end{equation}
A subtle distinction can be seen from the poles between the topological and nontopological regions. The relevant poles are $z_{0}, z_{1}^{+},$ and $z_{2}^{+}$. Within the topological region the $+$-poles lift off the real axis and become complex. Outside of this region the relevant poles remain all real, this distinction is shown in Fig.~\ref{fig:Poles}.

From this point, we can make a brief analysis of this expression to understand further how this simplifies for the two phases of the quantum model i.e. $\abs{\mu}<2w$ or $\abs{\mu}>2w$. First assume that $\zeta\in \mathbb{R}$. Then we have four cases:
\begin{enumerate}[label=(\roman*)]
	\item If $\abs{w\zeta \pm \Delta\mu}>0$
			\[ \abs{w\zeta + \Delta\mu}-\abs{w\zeta - \Delta\mu} = 2\mu\Delta \]
	\item If $\abs{w\zeta \pm \Delta\mu}<0$
			\[ \abs{w\zeta + \Delta\mu}-\abs{w\zeta - \Delta\mu} = -2\mu\Delta \]
	\item If $\abs{w\zeta + \Delta\mu}>0$ \& $\abs{w\zeta - \Delta\mu} <0$
			\[ \abs{w\zeta + \Delta\mu}-\abs{w\zeta - \Delta\mu} = 2w\zeta \]
	\item If $\abs{w\zeta + \Delta\mu}<0$ \& $\abs{w\zeta - \Delta\mu} >0$
			\[ \abs{w\zeta + \Delta\mu}-\abs{w\zeta - \Delta\mu} = -2w\zeta \]
\end{enumerate}
These lead to the four corresponding cases for the gap integral:
\begin{equation}
	\mathcal{E}_{gap} =
	\begin{cases}
		\frac{4\pi\epsilon\Delta^{2}}{w^{2}-\Delta^{2}}\left(\frac{\mu}{\zeta}-1\right), &\text{for case } (i)\\
		\frac{4\pi\epsilon\Delta^{2}}{w^{2}-\Delta^{2}}\left(-\frac{\mu}{\zeta}-1\right), &\text{for case } (ii)\\
		\frac{4\pi\epsilon\Delta^{2}}{w^{2}-\Delta^{2}}\left( -1+\frac{w}{\Delta} \right) = \frac{4\Delta}{w+\Delta}, &\text{for case }(iii)\\
		\frac{4\pi\epsilon\Delta^{2}}{w^{2}-\Delta^{2}}\left( -1 -\frac{w}{\Delta} \right) = \frac{-4\Delta}{w-\Delta}, &\text{for case }(iv).
	\end{cases}
\end{equation}
Finally, how do these cases correspond to the actual phase transition at $\abs{\mu}=2w$? Consider how we have split the case by the quantity $w\zeta\pm\mu\Delta$. This quantity changes sign when it crosses zero i.e. when
\begin{equation}
	\begin{split}
			w\zeta\pm\mu\Delta = 0,\\
			w^2(\mu^{2}-4w^{2}+4\Delta^{2}) = \mu^{2}\Delta^{2},\\
			\mu^{2}(w^{2}-\Delta^{2}) = 4w^{2}(w^{2}-\Delta^{2}),\\
			\Rightarrow \abs{\mu} = 2w,
	\end{split}
\end{equation}
provided that $w^{2}\neq \Delta^{2}$, which is the case throughout this work.

\end{document}